\documentclass[aps,prl,superscriptaddress,amsmath,amssymb,twocolumn,amsfonts]{revtex4-2}
\usepackage{bm,xcolor,MnSymbol}
\usepackage{graphicx}
\graphicspath{{Pictures/}}
\usepackage{braket,natbib}  
\usepackage[colorlinks=true,linkcolor=magenta,urlcolor=purple,citecolor=magenta,anchorcolor=blue]{hyperref}
\usepackage[normalem]{ulem}
\usepackage{orcidlink}
\usepackage{cancel}

\newcommand{\dg}{\ensuremath{\mathsf{g}}}
\newcommand{\vq}{\ensuremath{\mathbf{q}}}
\newcommand{\dr}{\ensuremath{\mathbf{r}}}

\newcommand{\bd}{\ensuremath{\mathcal{D}}}
\newcommand{\bs}{\ensuremath{\mathcal{S}}}
\newcommand{\cn}{\centering}

\newcommand{\SB}[1] {{\color{black}#1}}

\begin{document}
%----------------

\title{Multipolar multiferroics in $4d^2/5d^2$ Mott insulators}

%-----------------------------
\author{Saikat Banerjee\,\orcidlink{0000-0002-3397-0308}}
\email{saikat.banerjee@rutgers.edu}
\affiliation{Theoretical Division, Los Alamos National Laboratory, Los Alamos, New Mexico 87545, USA}
\affiliation{Center for Materials Theory, Rutgers University, Piscataway, New Jersey, 08854, USA}
\author{Stephan Humeniuk\,\orcidlink{0000-0002-7414-8468}}
\affiliation{Center for Materials Theory, Rutgers University, Piscataway, New Jersey, 08854, USA}
\author{Alan R. Bishop}
\affiliation{Theoretical Division, Los Alamos National Laboratory, Los Alamos, New Mexico 87545, USA}
\author{Avadh Saxena\,\orcidlink{0000-0002-3374-3236}}
\affiliation{Theoretical Division, Los Alamos National Laboratory, Los Alamos, New Mexico 87545, USA}
\author{Alexander V. Balatsky\,\orcidlink{0000-0003-4984-889X}}
\affiliation{Nordita, KTH Royal Institute of Technology and Stockholm University, 106 91 Stockholm, Sweden}
\affiliation{Department of Physics, University of Connecticut, Storrs, Connecticut 06269, USA}

\date{\today}

%-------------- Abstract --------------%
\begin{abstract}
\SB{We extend the concept of conventional multiferroicity \--- where ferroelectric and ferromagnetic orders coexist \--- to include multipolar degrees of freedom. Specifically, we explore how this phenomenon emerges in $4d^2/5d^2$ Mott insulators with strong spin-orbit and Hund’s couplings. Our study uncovers the origin of magnetic multipolar interactions in these systems and demonstrates that a combination of quadrupolar and octupolar magnetic order can simultaneously induce both electrical quadrupolar moments and ferroelectric polarization. By expanding the multiferroic framework to higher-order multipoles, we reveal the possibility of coexisting multipolar orders of different or same ranks, paving the way for new functional properties in a large class of strongly correlated materials.}
\end{abstract}
%-------------- Abstract --------------%

\maketitle
%-----------------------------

\textit{Introduction.}\textbf{\---} Materials that exhibit \SB{coexisting} magnetic and ferroelectric order, commonly known as multiferroics, open up a promising avenue for the mutual control of both magnetism and ferroelectricity. Although the research in multiferroics dates back to the 1950s~\cite{Landau_1984}, the recent advancements in theory, synthesis, and characterization techniques have significantly revived interest in these materials. Many candidate materials have been experimentally identified to harbor \SB{such} coexisting \SB{dipolar} orders~\cite{PhysRevB.94.201114,Fernandes2019}. Notably, there is growing experimental evidence for materials exhibiting even multipolar (quadrupolar, octupolar, etc.) or the so-called hidden orders (HO)~\cite{Patri2019,PhysRevResearch.2.022063,RevModPhys.81.807,Aeppli2020}. \SB{A compelling example is the long-standing puzzle of the HO phase transitions in URu$_2$Si$_2$~\cite{Mydosh2014}. This intriguing phenomenon is linked to the presence of magnetic multipolar moments, which also interact with various electrical multipolar moments, rendering it a rich arena for exploration and understanding in condensed matter physics~\cite{Haule2009,PhysRevLett.74.4301,Kasuya1997,PhysRevLett.81.3723,Chandra2002,PhysRevB.72.014432,PhysRevLett.106.106601}.} Therefore, it is natural to seek a microscopic description for coexisting multipolar orders.

\SB{Traditional multiferroic research~\cite{Mangalam2009, Evans2015, Belik_2011, Lin2017,Tokura2000,Tokura2010,Tokura2019} has mainly explored how dipolar limits of spin and charge degrees-of-freedom (DF) behave in different materials, including metals, semiconductors, and Mott insulators (MIs)~\cite{Fiebig_2005, Eerenstein2006, Ramesh2007, Cheong2007, Khomskii_2009, Dong2019, LottermoserMeier_2021}. In simple MIs, the electric charge is \SB{energetically} gapped out, leaving only spins to determine the low-energy properties of the material. The arrangement of these spins creates various magnetic orders.

In certain MIs with strong spin-orbit coupling, a non-collinear spin pattern can break inversion symmetry, leading to ferroelectric polarization, known as the Katsura-Nagaosa-Balatsky (KNB) mechanism~\cite{PhysRevLett.95.057205, Tokura_2014, Aeppli2020, Bossini_2023}. Conversely, in some mixed-valence compounds (improper ferroelectrics), pre-existing ferroelectric polarization can generate magnetization~\cite{PhysRevMaterials.1.071401,Ederer2004}. Interestingly, even magnetic systems that lack conventional order can still exhibit both magnetization and electric polarization, revealing new types of multiferroic behavior~\cite{PhysRevB.78.024402, Banerjee2023A, Banerjee2023B}.}

The complexity arises when a MI lacks any dipolar (spin) DF but is only governed by higher-order multipoles \--- a common scenario in heavy fermionic or \SB{$4d$/5$d$} transition metal (TM) compounds due to a subtle interplay of crystal field effects, Hund's coupling, and \SB{a large} spin-orbit coupling (SOC)~\cite{Krempa2014,Takayama2021}. Here dipolar \SB{order} is absent; leading to first non-vanishing moment as quadrupole or octupole moment, etc.~\cite{PhysRevB.84.094420,PhysRevResearch.3.033163,PhysRevB.107.L020408,PhysRevB.101.054439,PhysRevLett.124.087206,PhysRevB.101.155118,PhysRevB.105.014438}. Although multipole contribution is typically weaker than the dipole order in common systems, its usefulness is of \SB{fundamental} importance if the dipole \SB{component} is intrinsically absent. This naturally leads to the question: can we expect an emergent ferroelectric polarization induced by such magnetic multipoles? \SB{More generally}, is there a potential scheme for cross-coupling between the magnetic and electric multipoles?

\SB{Here, we demonstrate that $d^2$ ions with strong SOC and Hund’s coupling can give rise to both ferroelectric polarization and electrical quadrupolar moments. We begin by examining the microscopic details of $d^2$ systems~\cite{PhysRevResearch.3.033163} and derive a multipolar exchange model relevant to edge-shared octahedral geometries. Using classical Monte Carlo (MC) simulations, we identify non-collinear magnetic octupolar textures (see Fig.~\ref{fig:Fig3}), which suggest the possibility of induced ferroelectric response. Furthermore, our framework predicts a ferroquadrupolar charge response that emerges purely from magnetic quadrupolar textures, even in the absence of non-collinearity \--- a key distinction from the conventional KNB-type multiferroicity.  

The central result of our work is the framework for ``multipolar multiferroicity", which extends the conventional multiferroic paradigm beyond dipole-dipole coupling. This insight opens up new avenues for exploring multipolar degrees of freedom and their novel functionalities in correlated materials. We also discuss potential experimental signatures and quantitative estimates to guide future studies.}

\textit{Multipolar multiferroic framework.}\textbf{\---} Here, we present a simplified mathematical framework for the possible multipolar multiferroic coupling in MIs. We adopt a convention in which the ``atomic" low-energy DF are characterized by the Stevens operators $\mathcal{O}_{lm}$~\cite{Kusunose2008}. The lowest-order Stevens operators are \SB{simply} the spin DF~\cite{Takayama2021}. \SB{As we will see, in complex systems such as $d^2$ Mott insulators, the low-energy behavior is governed by higher-order Stevens operators, resulting in an effective multipolar exchange model. These higher-order interactions promote the emergence of diverse multipolar orders, including quadrupolar and octupolar states, in the parent Mott insulators}~\cite{PhysRevB.101.054439,PhysRevB.101.155118,PhysRevLett.124.087206,PhysRevB.94.201114,PhysRevResearch.3.033163,PhysRevB.107.L020408,PhysRevLett.127.237201}.

%-----------------------------------------------
\begin{table}[t]
\centering
\caption{A summary of the higher-order multiferroic couplings. Notation: $l = 1$ $\rightarrow$ dipolar, $l = 2$ $\rightarrow$ quadrupolar, $l = 3$ $\rightarrow$ octupolar etc., and $l_{\rm{E(M)}} \rightarrow$ electrical (magnetic) sectors. The hetero-(homo)-geneous multiferroicity signifies coupling between different (same) order multipoles.}\label{tab:multi_ferroic}
\begin{tabular}{|p{1cm}|p{1cm}|p{2.2cm}|p{1cm}|p{1cm}|}
\hline \hline
\multicolumn{2}{|c|}{Conventional} & \multicolumn{3}{|c|}{Multipolar} \\
\hline \hline
\cn $l_{\rm{E}}$ & \cn $l_{\rm{M}}$ & \cn Type & \cn $l_{\rm{E}}$ & \; \; $l_{\rm{M}}$ \\
\hline
\cn -	&	\cn -	&	\cn Heterogeneous & \cn 1	& \; \;  $3$	\\
\hline
\cn 1 & \cn 1	 &	\cn Homogeneous & \cn 2 	& \; \;  $2$	\\
\hline
\multicolumn{2}{|c|}{Ref.~\cite{PhysRevLett.95.057205,Cheong2007}} & \multicolumn{3}{|c|}{This work} \\
\hline \hline
\end{tabular}
\end{table}
%-----------------------------------------------
Since the charge DF are gapped within a MI, the electric multipole moments can be induced only via the virtual exchange processes of the electrons \SB{with the ligand sites} in the absence of doping. For a discrete charge distribution $\rho(\dr_i)$, such multipole moments are defined as
%-----------------------------
\begin{equation}\label{eq.1}
\mathcal{Q}_{lm} 
= 
e \sum_{i=1}^{N} \int Z^{\ast}_{lm}(\hat{\delta\dr}_i) \delta\dr_i^{l} \psi^\ast_{\rm{f}}(\delta\dr_i) \psi_{\rm{i}}(\delta\dr_i) d^3\dr,
\end{equation}
%-----------------------------
where $Z_{lm}(\hat{\delta\dr}_i) = \sqrt{\tfrac{4\pi}{2l+1}} Y_{lm}(\hat{\delta\dr}_i)$ are the spherical harmonics with Racah normalization, $\delta\dr_i = \dr -\dr_i$, $N$ is the number of sites in a finite-size cluster, and $\psi_{\rm{f,i}}$ are the initial and final states \SB{during the virtual electron exchange processes}. Subsequently, we adopt the real representation of $\mathcal{Q}_{lm}$ by constructing the tesseral harmonics as~\cite{Kusunose2008,Hayami2018}
%-----------------------------
\begin{equation}\label{eq.2}
\mathcal{Q}^{{\rm{c}}}_{lm} = (-1)^m \tfrac{\mathcal{Q}_{lm} + \mathcal{Q}^{\ast}_{lm}}{\sqrt{2}}, \quad
\mathcal{Q}^{{\rm{s}}}_{lm} = (-1)^m \tfrac{\mathcal{Q}_{lm} - \mathcal{Q}^{\ast}_{lm}}{\sqrt{2}i}.
\end{equation}
%-----------------------------
Rewriting Eq.~\eqref{eq.1} in this way is beneficial as both the electric and magnetic multipolar moments transform identically under the point-group representations of the underlying crystal. Under two discrete symmetries: parity/inversion ($\mathcal{P}$) and time-reversal ($\mathcal{T}$), the corresponding multipolar moments transform under these two symmetries as $\mathcal{P}: \mathcal{Q}^{\rm{c/s}}_{lm} \rightarrow (-1)^{l} \mathcal{Q}^{\rm{c/s}}_{lm}$, and $\mathcal{T}: \mathcal{O}_{lm} \rightarrow (-1)^{l} \mathcal{O}_{lm}$. Since they transform identically, the obvious possibility is whether a specific configuration of magnetic multipolar moments can trigger multipolar charge responses and vice versa. Our prediction of magnetic  octupole-induced ferroelectricity and magnetic quadrupole-induced electrical quadrupolar responses exemplifies this concept and motivates us to propose an extended classification of multiferroics beyond conventional frameworks, as illustrated in Table~\ref{tab:multi_ferroic}.      

%-------------------------------------------------------------------------------------------
\begin{figure}[t!]
\centering \includegraphics[width=\columnwidth]{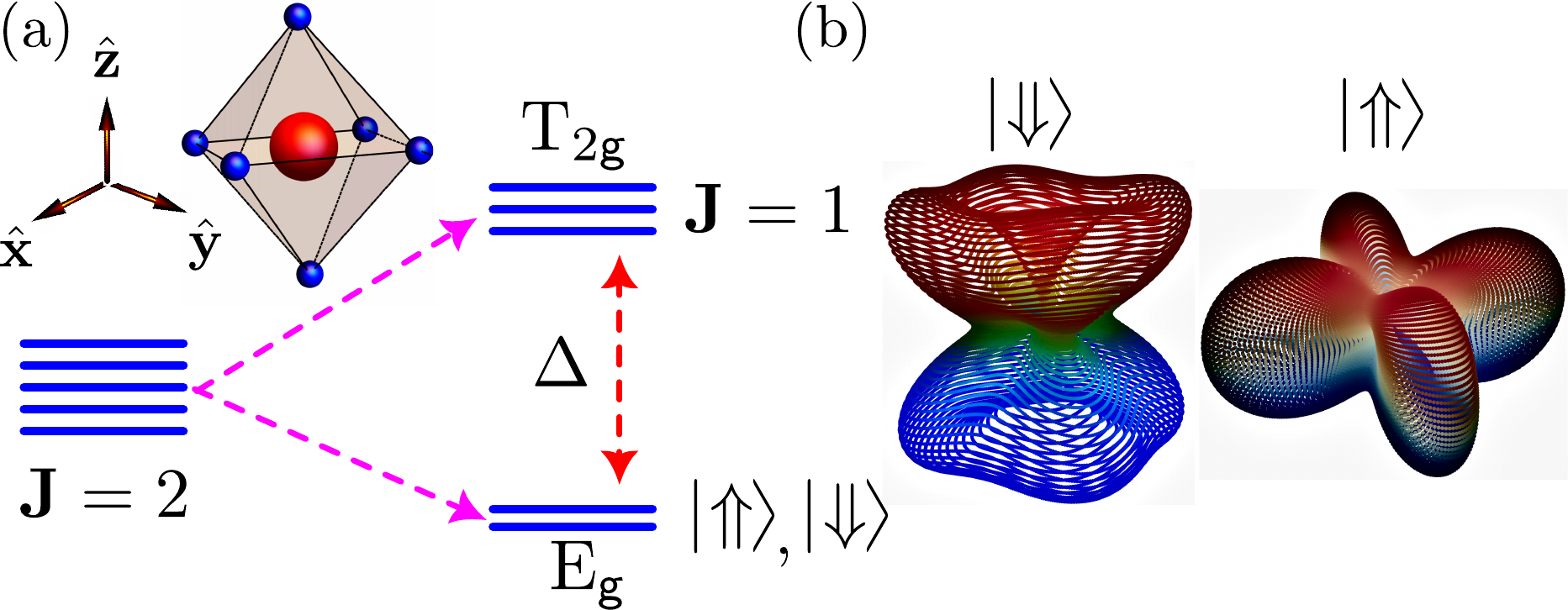}
\caption{(a) Left: Schematic of an octahedral unit-cell with TM atoms (red sphere) surrounded by the ligand (blue spheres). The crystal field further splits the ${\bf{J}} = 2$ states into a $\rm{E}_{\dg}$ doublet and a $\rm{T}_{2\dg}$ triplet. Right: The corresponding $\rm{T}_{2\dg}$, and $\rm{E}_{\dg}$ splitting of ${\bf{J}} = 2$ atomic energy levels under the influence of crystal field ($\Delta_{\rm{crys.}}$), Hund's coupling ($J_{\rm{H}}$) and SOC ($\lambda$). Here, $\Delta$ denotes a renormalized crystal field acting on the ${\bf{J}} = 2$ manifold. (b) The spatial distribution of the charge density for the pseudo-spin up and down states.}\label{fig:Fig1}
\end{figure}
%-------------------------------------------------------------------------------------------

\textit{Model.}\textbf{\---} \SB{We begin with} a $d^2$ ion in an octahedral crystal field, \SB{where} the atomic ground state is the non-Kramers $\rm{E}_{\sf{g}}$ doublet characterized by the total angular momentum operator $\bf{J} = \bf{L} + \bf{S}$~\cite{PhysRevResearch.3.033163}. The corresponding atomic energy splitting is illustrated in Fig.~\ref{fig:Fig1}, with the $\rm{E}_{\sf{g}}$ states given by
%----------------------------
\begin{equation}\label{eq.3}
\ket{\Uparrow} = \frac{1}{\sqrt{2}} (\ket{J_z = 2} + \ket{J_z = -2}), \quad
\ket{\Downarrow} = \ket{J_z = 0}.
\end{equation}
%----------------------------
The detailed spatial structure of these states with two spin-orbit coupled $d$-electrons in the crystal field splitting is given in the Supplementary Material (SM)~\cite{supp}. The real-space charge distribution for the pseudo-spin states is shown in Fig.~\ref{fig:Fig1}(b). We notice that the $\ket{\Uparrow}$-state is distributed along the $xy$-plane since one of the electrons has to occupy a $d_{xy}$ orbital, while the $\ket{\Downarrow}$-state is elongated along the $z$-axis as two electrons can simultaneously occupy the $d_{yz}$ and $d_{zx}$ orbitals~\cite{supp}.  \SB{It is easy to show that the eigenstates in Eq.~\eqref{eq.3} host non-zero matrix elements between} three higher-order magnetic Stevens operators, namely $\mathcal{O}_{20} = 3J_z^2-{\bf{J}}^2$, $\mathcal{O}_{22} = J_x^2 - J_y^2$, and ${\sf{T}}_{xyz} = \overline{J_x J_y J_z}$ (average corresponds to symmetrization over all indices)~\cite{PhysRevB.84.094420,Patri2019,PhysRevResearch.3.033163,PhysRevB.101.054439,PhysRevLett.124.087206}. Normalizing the operators as follows
%-----------------------------
\begin{equation}\label{eq.4}
\frac{\mathcal{O}_{22}}{4\sqrt{3}} \rightarrow \tilde{\sigma}^x, 	\qquad
\frac{\mathsf{T}_{xyz}}{2\sqrt{3}} \rightarrow \tilde{\sigma}^y, 	\qquad
\frac{\mathcal{O}_{20}}{12} \rightarrow \tilde{\sigma}^z,
\end{equation}
%-----------------------------
it can be shown that they satisfy an effective SU(2) pseudo-spin algebra, \textit{i.e.} $[\tilde{\sigma}^{\alpha}, \tilde{\sigma}^{\beta}] = i \epsilon_{\alpha\beta\gamma} \tilde{\sigma}^{\gamma}$. 

At first, we note that under time-reversal symmetry (TRS) transformation, pseudo-spin operators transform as ${\mathcal{T}}: \{ \tilde{\sigma}^x, \tilde{\sigma}^y, \tilde{\sigma}^z \} \rightarrow \{ \tilde{\sigma}^x, - \tilde{\sigma}^y, \tilde{\sigma}^z \}$,  as evident from Eq.~\eqref{eq.4}. This fact constrains the possible bilinear and trilinear exchange terms \SB{in the governing Hamiltonian}. For example, terms including a single $\tilde{\sigma}^y$ like $\tilde{\sigma}^{\eta}_i\tilde{\sigma}^y_j (\eta = x,z)$ are not allowed in the bilinear exchange. Furthermore, the trilinear exchange may contain terms like $\tilde{\sigma}^y_i\tilde{\sigma}^y_j \tilde{\sigma}^{x/z}_k$ ($\forall$ $i,j,k$ neighboring sites), and any other combinations. However, such terms would be forbidden by $\sf{C}_3^{[111]}$ symmetry, \SB{as} $\tilde{\sigma}^x$, $\tilde{\sigma}^z$ transform like a conventional $e_{\dg}$ orbital in a Kugel-Khomskii model~\cite{Kugel_1982,Abragam_1970,Khaliullin_2005}, while $\tilde{\sigma}^y$ remains invariant. \SB{In general, the explicit form of the multipolar Hamiltonian can be complex, as it depends on the specific material properties. However, such an exchange Hamiltonian can, in principle, give rise to non-trivial multipolar textures, as we show below.}

\begin{figure}[t!]
\centering \includegraphics[width=\columnwidth]{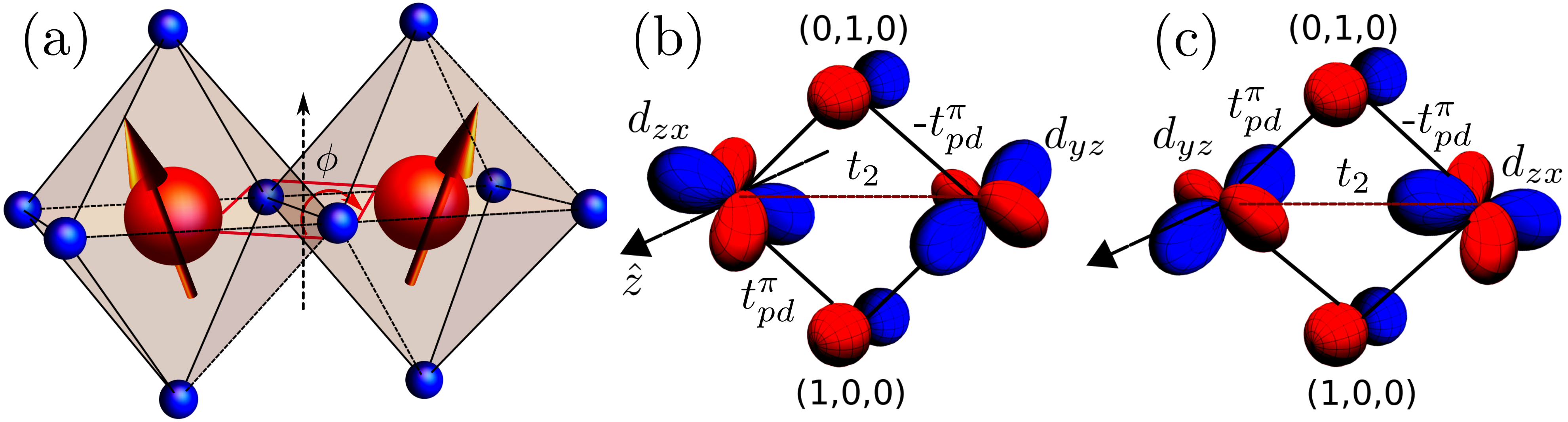}
\caption{The different hopping channels to the ligand $p$ orbitals in a four-site cluster. (a) A double-exchange hopping from the ligand to the TM sites, and (b) a cotunneling of electrons (white filled circles) from the TM to the two ligand sites, relevant for ferroelectric polarization, and electrical quadrupolar moments, respectively. Other path corresponds to the remaining possible hopping in the left-over channel.}\label{fig:Fig2}
\end{figure}

\begin{figure}
\centering
\includegraphics[width=0.9\linewidth]{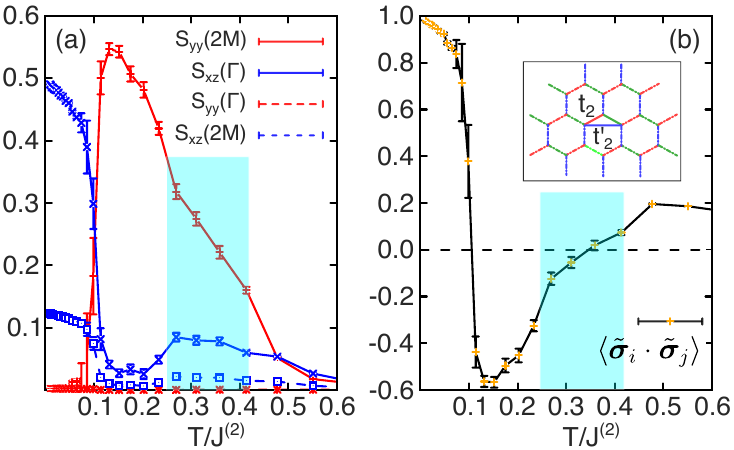}
\caption{(a) Structure factor of the octupolar ($S_{yy}$) and quadrupolar ($S_{xz}$) spin components at momentum points $\Gamma$ and $2M$. The strength of the three-spin term is $\zeta = 0.15$ and the system size is $12 \times 12 \times 2$. For $T/J^{(2)} < 0.1$, the order is quadrupolar ($xz$) FM, whereas for $0.1 < T/J^{(2)} < 0.5$ it is octupolar ($y$) AFM. In the shaded region, the spin configuration is canted, i.e., ($y$) AFM and ($xz$) FM orders coexist. (b) Average \SB{scalar} product between neighboring spins as a crude measure of collinearity. \SB{(Inset) A cartoon of the honeycomb structure with relevant hoppings as in Eq.~\eqref{eq.8}. Different colors label different bond-types.}}\label{fig:Fig3}
\end{figure}

\textit{Ferroelectric polarization.}\textbf{\---} \SB{Without adhering to any specific scenario as mentioned earlier}, here we take a local approximation of a \SB{generic} multipolar Hamiltonian as $\mathcal{H}_{\rm{local}} = - \tilde{U} \sum_{i} \hat{\bm{e}}_i \cdot \tilde{\bm{\sigma}}_i$, where $\tilde{U}$ denotes the local excitation energy, and $\hat{\bm{e}}_i = (\cos \phi_i \sin \theta_i, \sin \phi_i \sin \theta_i, \cos \theta_i)$ describes \SB{the local orientation of the} pseudo-spin moments at site $i$~\cite{PhysRevLett.95.057205}. Next, \SB{for the virtual charge transfer process}, we consider a four-site cluster of two TM and two ligand atoms in the edge-shared geometry as shown in Fig.~\ref{fig:Fig2}(a). \SB{We focus on the stoichiometric configuration with electron \textit{filled} ligand sites}. Subsequently, we allow ligand-TM charge transfer processes in a double-exchange (DE) framework and consider two different hopping paths ($t^{\pi}_{pd} > 0$) in the upper and lower triangular plaquettes as in Fig.~\ref{fig:Fig2}(a). The eigenstates are obtained by diagonalizing $\mathcal{H}_{\rm{local}}$ as follows
%--------------------------
\begin{equation}\label{eq.5}
\begin{split}
\ket{\psi^{+}} & = \cos \tfrac{\theta_i}{2} \ket{\Uparrow} + e^{i\phi_i} \sin \tfrac{\theta_i}{2} \ket{\Downarrow}, \\
\ket{\psi^{-}}  & = \sin \tfrac{\theta_i}{2} \ket{\Uparrow} - e^{i\phi_i} \cos \tfrac{\theta_i}{2} \ket{\Downarrow},
\end{split}
\end{equation}
%--------------------------
where $\ket{\Uparrow},\ket{\Downarrow}$ are the same as in Eq.~\eqref{eq.3}, with energies $\mp\tilde{U}$. Assuming a finite TM-ligand hopping ($t^\pi_{pd}$), we compute the perturbation correction to the ground states in Eq.~\eqref{eq.5} leading to an intermediate configuration: $d^3,p_z^1$, where a strong Hund's coupling is considered for the $d^3$ ion in the octahedral crystal field. Utilizing these perturbed states $\ket{\psi}$ in the four-site cluster~\cite{supp}, we analyze the expectation value of the ferroelectric polarization ${\bf{P}} = \braket{\psi|e\dr|\psi}$. We notice that for the $z$-bond, only $x$-component of $\bf{P}$ is finite and is given by 
%------------------------------
\begin{equation}\label{eq.6}
\braket{P_x} \approx \frac{eIt^{\pi}_{pd}}{3\Delta_{\rm{crys.}}} 
\cos \frac{\theta_i}{2} \cos \frac{\theta_j}{2} \sin (\phi_i - \phi_j ) \,, 
\end{equation}
%------------------------------
where the overlap integral $I$ is defined as  $I \sim\int d\dr \; x p_{\dr}^z d_{\dr}^{zx}$ and its cyclic permutations. \SB{The coordinate system is centered at the middle of the four-site cluster and aligned with the local octahedral coordinate axes.} The emergent polarization is \SB{manifestly} odd under inversion symmetry and is bond-dependent. For other bonds, the finite component is obtained by cyclic permutation of $x,y,z$. \SB{We note that} without any octupolar component, \textit{i.e.} $\phi_i = 0$, the polarization vanishes \--- aligning with our earlier phenomenological analysis (see Table~\ref{tab:multi_ferroic}, heterogeneous type: $l_{\rm{E}}=1, l_{\rm{M}}=3$). Even a coplanar quadrupole order along with a non-collinear octupole part leads to a finite ferroelectric polarization. \SB{Unlike the KNB mechanism~\cite{PhysRevLett.95.057205}, this mechanism can generate a finite ferroelectric polarization without sharp features in neutron scattering, indicating absence of spin moments in materials with multipolar order. Approximating the orbitals by hydrogen-like wavefunctions, it is found that the overlap integral $I$ is an order of magnitude larger for $4d$ and $5d$ orbitals compared to $3d$ orbitals. With typical lattice constants a rough estimate of the polarization is then \mbox{$\langle P_x \rangle \sim  (\tfrac{t_{pd}^{\pi}}{\Delta_{\text{crys.}}}) \mu {\rm{C}}\; \rm{cm}^{-2}$}, similar to or even larger than the estimate in Ref.~\cite{PhysRevLett.95.057205} for $3d$ orbitals.}

%-------------------------------------------------------------------------------------------
%-------------------------------------------------------------------------------------------

\textit{Electrical quadrupole moment.}\textbf{\---} \SB{We now turn to ferroquadrupolar charge distribution} on the same geometry, \SB{however, focused on the stoichiometric configuration with electron \textit{empty} ligand sites}. A cotunneling \SB{process is considered}, where each TM electron simultaneously \SB{hops to upper and lower} empty ligand sites (\SB{such as boron group of elements}), as illustrated in Fig.~\ref{fig:Fig2}(b). \SB{This can be a weak but a physical process}. We employ a similar perturbative approach and focus on the correction to the ground state in Eq.~\eqref{eq.5} leading to the intermediate configuration: $d^0,p_z^1,p_z^1$. Again utilizing the perturbed states $\ket{\psi}$ within the four-site cluster~\cite{supp}, we analyze the expectation value of the electrical multipole moment $\mathcal{Q}_{lm}$ as defined in Eq.~\eqref{eq.2}. In this case, we find that for the $z$-bond only $\mathcal{Q}^{\rm{s}}_{21}$ and $\mathcal{Q}^{\rm{s}}_{22}$ are non-vanishing. Setting the origin at the center of the cluster, we obtain
%------------------------------
\begin{subequations}
\begin{align}
\label{eq.7.1}
& \mathcal{Q}^{\rm{s}}_{21}
\approx
\frac{8e I' (t^{\pi}_{pd})^2}{3\sqrt{3}\Delta_{\rm{crys.}}^2} 
\cos \frac{\theta_i}{2} \cos \frac{\theta_j}{2} \cos (\phi_i - \phi_j), \\
\label{eq.7.2}
\mathcal{Q}^{\rm{s}}_{22} 
\approx &
\frac{4e I' (t^{\pi}_{pd})^2}{3\Delta_{\rm{crys.}}^2} 
\cos \frac{\theta_i}{2} \sin \frac{\theta_j}{2} \cos \phi_i
+ i \leftrightarrow j + \mathcal{Q}^{\rm{s}}_{21},
\end{align}
\end{subequations}
%------------------------------
where the overlap integral is defined as $I' \sim \int d\dr d_{\dr}^{xy} d_{\dr}^{zx} yz p_{\dr}^z p_{\dr}^{z} \approx I^2$, and its cyclic permutations. Here, $l$ $(l')$ corresponds to the upper (lower) ligand sites in Fig.~\ref{fig:Fig2}(b). For other bonds, the non-vanishing quadrupolar components are obtained by cyclic permutations of $x,y,z$. Note that, even without any magnetic octupolar moments, the electrical quadrupolar moment survives and arises only from the magnetic quadrupole, consistent with Table~\ref{tab:multi_ferroic}, homogeneous type: $l_{\rm{E}}=2, l_{\rm{M}}=2$. \SB{To the best of our knowledge, this stands in sharp contrast to all known multiferroic mechanisms to date, and indeed provides a route to multipolar generalization. Taking a similar approach as before, the induced quadrupolar magnitude (in the unit of $\mu \rm{C}\, cm^{-1}$) will be relatively small compared to the ferroelectric polarization.}

\SB{Having obtained the multipolar response functions, we now explore} the effective multipolar exchange interaction by performing perturbation theory on the Hubbard-Kanamori model. We consider an ideal edge-sharing octahedral geometry, and focus on the [111] plane, where the TM sites form a honeycomb lattice~\cite{PhysRevResearch.3.033163} (see inset in Fig.~\ref{fig:Fig3}(b)). The microscopic Hamiltonian is given by $\mathcal{H} = \mathcal{H}_0 + \mathcal{H}_1$, where 
%----------------
\begin{align}
\nonumber
& \mathcal{H}_0 = 
U\sum_{i\alpha} n_{i\alpha\uparrow}n_{i\alpha\downarrow} + 
\frac{U'}{2}\sum_{\substack{i,\alpha\neq \beta \\ \sigma\sigma'}} n_{i\alpha \sigma} n_{i\beta \sigma'} +  \Delta_{\rm{crys.}} \sum_{l,\sigma}n^p_{l\sigma} +\\
\nonumber
& -\frac{J_\mathrm{H}}{2}\sum_{\substack{i,\alpha\neq \beta \\ \sigma \sigma'}} d^{\dagger}_{i\alpha \sigma} d_{i\alpha\sigma'} d^{\dagger}_{i\beta \sigma'}  d_{i\beta \sigma}^{\phantom\dagger} 
+ \lambda \sum_{\substack{i,\alpha \beta \\ \sigma \sigma'}} d^{\dagger}_{i\alpha \sigma} \left( \mathbf{L} \cdot \mathbf{S} \right)_{\substack{\alpha \beta \\ \sigma \sigma'}} d_{i \beta \sigma'} , \\
\label{eq.8}
&\mathcal{H}_1 
= 
\sum_{\substack{ij, \eta \\ \alpha \beta}} t^{(\eta)}_{\alpha\beta} d^{\dag}_{i \alpha \sigma} d_{j \beta \sigma'} + 
t^{(\eta)}_{pd} \sum_{\substack{\langle il \rangle, \eta \\ \alpha \sigma}} p^{\dag}_{l \sigma} d_{i\alpha \sigma} +  \mathrm{h.c.},
\end{align}
%----------------
where $U$, $U'$ denote the intra- and inter-orbital Coulomb repulsion, $J_{\rm{H}}$ is the Hund’s coupling, $\Delta_{\rm{crys.}}$ denotes the ligand charge-transfer energy, and $\lambda$ is the strength of the SOC, and $t^{(\eta)}_{\alpha\beta},t'^{(\eta)}_{\alpha\beta}$ ($t^{(\eta)}_{pd}$) correspond to the various bond-dependent hopping amplitudes as shown in Fig.~\ref{fig:Fig2}(b). For simplicity, we integrate out the ligand degrees of freedom and consider a renormalized TM-TM hopping as $t^{(\eta)}_{\alpha\beta} \rightarrow t^{(\eta)}_{\alpha\beta} + (t^{(\eta)}_{pd})^2/\Delta_{\rm{crys.}}$~\cite{PhysRevB.98.125135}, with $t^{(\eta)}_{\alpha\beta}$ obtained via Slater-Koster integrals~\cite{PhysRev.94.1498}. Projecting Eq.~\eqref{eq.5} into the $\rm{E}_\dg$ doublet, we obtain the exchange Hamiltonian as $\mathcal{H} = \sum_\eta \mathcal{H}_\eta + \mathcal{H}^{(3)}$ where~\cite{supp}
%--------------------
\begin{equation}\label{eq.9}
\mathcal{H}_{\eta} = J^{(2)}
\sum_{\langle ij \rangle, \eta} 
\left(  
\tilde{\sigma}^y_{i} \tilde{\sigma}^y_{j} 
- 
\tilde{\sigma}^x_{i} \tilde{\sigma}^x_{j} - 
\tilde{\sigma}^z_{i} \tilde{\sigma}^z_{j} 
\right),
\end{equation}
%--------------------
and the bilinear exchange interaction is identical along all the bond directions $\eta \in \{ x,y,z\}$ consistent with the cubic rotation symmetry of the pseudo-spin DF~\cite{PhysRevResearch.3.033163}, and $J^{(2)} = 2t_2^2/(3 \Delta E_{\rm{ex.}})$, where $\Delta E_{\rm{ex.}} = U - 3 J_{\rm{H}} + \lambda$. The expression in Eq.~\eqref{eq.9} may give the misleading impression that the model remains valid even in the absence of $J_{\rm{H}},\lambda$. However, \SB{it is true only} in the presence of the non-Kramers doublet induced by Hund's coupling and SOC [see Fig.~\ref{fig:Fig1}(a)]. This validity is further corroborated by numerical exact diagonalization calculations performed on a two-site cluster, as reported in Ref.~\cite{PhysRevB.104.174431}. \SB{The trilinear exchange term is inherently complex and becomes finite only when TRS is broken by an external field. Its explicit form is obtained through the third-order perturbation theory and, as discussed earlier, is constrained by both TRS and $\sf{C}_3$ symmetry. The latter reads as~\cite{supp}} 
%-----------------------------
\begin{equation}\label{eq.10}
\mathcal{H}^{(3)} = \chi_s
\sum_{\llangle ijk \rrangle}
 \left(
\alpha\tilde{\sigma}_{i}^{x}\tilde{\sigma}_{j}^{x} + \alpha \tilde{\sigma}_{i}^{z}\tilde{\sigma}_{j}^{z} -
\beta \tilde{\sigma}_i^{y} \tilde{\sigma}_j^{y} \right) \tilde{\sigma}_k^{y} + P_{ijk}, \\
\end{equation}
%-----------------------------
where $\llangle ijk \rrangle$ denotes all possible triangles of nearest-neighbor sites on the honeycomb lattice, $P_{ijk}$ indicates the cyclic permutations of the site indices $i,j,k$, $\chi_s = 2t_2^2 t_2'/(3\Delta E_{\rm{ex.}}^2) \sin{\tfrac{e\Phi}{\hbar c}}$, with $\Phi$ being the total flux in the triangular plaquette in an external magnetic field $B$~\cite{Diptiman_PRB1995}, and $\alpha=\tfrac{2}{9}, \beta=\tfrac{1}{9}$~\cite{supp}. Here, we ignored the virtual hopping between $\rm{T}_{2\dg}$ and $\rm{E}_\dg$ sectors, and ignored the Zeeman coupling. However, we assumed the following hierarchy $U > \Delta_{\rm{crys.}} > J_{\rm{H}},\lambda > B$ in the energy scales. The effect of \SB{	a larger} Zeeman coupling is known to induce quadrupole-octupole coupling as reported in a previous theoretical work~\cite{PhysRevB.107.L020408}.  

%----------------------------------
%----------------------------------

\textit{Multipolar ordering.}\textbf{\---} \SB{Having derived the exchange Hamiltonian in Eqs.~\eqref{eq.9} and \eqref{eq.10}, we now utilize classical MC simulations to explore potential pseudo-spin textures in an external magnetic field, consistent with our local approximations used in the derivation of Eqs.~\eqref{eq.6}, \eqref{eq.7.1}, and \eqref{eq.7.2}.} In this regard, we tune the relative strength of the bilinear and trilinear terms in the Hamiltonian. We label this parameter as $\zeta = \chi_s / J^{(2)} = t_2'/\Delta E_{\rm{ex.}} \sin \tfrac{e\Phi}{\hbar c}$. 

We adopt a classical approximation as $\tilde{\bm{\sigma}}_i = (\sin \theta_i \cos \phi_i, \sin \theta_i \sin \phi_i, \cos \theta_i)$, where each pseudo-spin is set to be an O(3) vector satisfying $|\tilde{\bm{\sigma}}_i \cdot \tilde{\bm{\sigma}}_i| = 1$. The simulations are performed using the standard Metropolis single-spin flip update \cite{Metropolis2004} and parallel tempering \cite{Hukushima1996} with a geometric temperature grid. Without the three-spin term ($\zeta = 0$), the Hamiltonian in Eq.~\eqref{eq.9} can be recast as the Heisenberg antiferromagnet by a gauge transformation $\tilde{\sigma}^{x,z}_i \rightarrow -\tilde{\sigma}^{x,z}_i$ on one of the two sublattices, and the ground state is a N\'{e}el antiferromagnet. This corresponds to antiferromagnetically (AFM) aligned octupolar ($y$) components and ferromagnetically (FM) aligned quadrupolar ($xz$) components of equal strength. 

As Fig.~\ref{fig:Fig3}(a) shows, the three-spin term Eq.~\eqref{eq.10} changes the balance between octupolar and quadrupolar components: The ground state and low-temperature phase is $xz$ FM, which is separated by a sharp spin-flop transition from a predominantly $y$ AFM phase. This is evidenced by the \SB{static} structure factor for the quadrupolar, $S_{xz}({\vq}) = \frac{1}{N^2} \sum_{ij} e^{i \vq \cdot (\dr_i - \dr_j)} \langle \frac{1}{2}(\tilde{\sigma}_i^{x} \tilde{\sigma}_j^{x} + \tilde{\sigma}_{i}^{z} \tilde{\sigma}_j^{z})\rangle$, and the octupolar, $S_{yy}(\vq) = \frac{1}{N^2} \sum_{ij} e^{i \vq \cdot (\dr_i - \dr_j)} \langle \tilde{\sigma}_i^{y} \tilde{\sigma}_j^{y} \rangle$ components: In the $xz$-FM phase $S_{xz}(\vq)$ has a peak at the center of the first Brillouin zone (BZ) $\Gamma = (0,0)$ and a second peak of $1/4$ its strength at the point $2{\rm{M}} = (\frac{2\pi}{a}, \frac{2\pi}{\sqrt{3}a})$ at the center of the second BZ, which indicates intra-unit-cell order. Here, $a$ is the lattice constant of the underlying triangular Bravais lattice as shown in Fig.~\ref{fig:Fig3}(b). By contrast, in the $y$-AFM phase $S_{y}(2\rm{M})$ is large, while $S_{yy}(\Gamma)$ vanishes. Importantly, there is a finite-temperature window where both $S_{yy}(2\rm{M})$ and $S_{xz}(\Gamma)$ are non-zero and of unequal strength, which indicates a non-collinear, canted spin configuration that is AFM in $y$-components and FM in $xz$-components. This region is shaded in Fig.~\ref{fig:Fig3}(a,b). 

As a crude measure of collinearity we consider the scalar product between nearest neighbor pseudo-spins, $\langle \tilde{\bm{\sigma}}_i \cdot \tilde{\bm{\sigma}}_{\langle j \rangle_i} \rangle$, averaged over all bonds $(i,\langle j \rangle_i)$, which is shown in Fig.~\ref{fig:Fig3}(b) and is consistent with a quadrupolar FM polarized state at low-temperatures and an octupolar AFM or canted state at intermediate temperatures. The implication of quantum or thermal fluctuations is an important question here but is beyond the scope of this work. \SB{This MC simulation for potential multipolar textures is an \textit{ad hoc} and crude proof for the underlying approximations adopted earlier for the multiferroic responses in Eq.~\eqref{eq.6}, and Eqs.~\eqref{eq.7.1} and \eqref{eq.7.2}.}

\textit{Experimental signatures}\textbf{\---} Our proposed functionality can be realized in $d^2/f^2$ octahedral ions \SB{such as pressurized ReCl$_5$~\cite{Mucker1968} (relevant for ferroelectric polarization), or [ReIn$_6$]$^{3-}$, and [Re(B$_6$H$_6$)]$^{3-}$~\cite{Palmer2015} (relevant for ferroquadrupolar response) metal complexes}. While detecting multipolar orders presents significant challenges, we outline several potential experimental probes. Magnetic octupolar order, for instance, can lead to transverse magnetic fields under shear strain \--- known as transverse piezomagnetism~\cite{Cheong2024a,Cheong2024b}. Similarly, magnetic quadrupolar order can couple to specific normal modes of Jahn-Teller distortions~\cite{Patri2019,PhysRevResearch.3.033163}, which can be precisely measured using x-ray diffraction experiments, thereby indirectly confirming the magnetic quadrupolar orders~\cite{PhysRevResearch.2.022063}. Optical techniques such as second-harmonic generation or the Kerr effect can be employed to detect the presence of induced ferroelectric polarization. Although detecting ferroquadrupolar moments is exceptionally challenging, nuclear magnetic resonance spectroscopy may provide a viable pathway for its observation. Ab-initio~\cite{PhysRevResearch.5.L012010}, or cluster dynamical mean-field theory calculations~\cite{PhysRevB.100.235117}, may provide \SB{more accurate and} realistic estimates.

\textit{Discussion and perspective.}\textbf{\---} In this Letter, we proposed a mechanism generalizing the conventional multiferroicity to higher-order sectors. \SB{Even if the multipolar magnitudes are small compared to their dipolar counterparts, identifying materials where dipolar electric and magnetic moments vanish is fundamentally important, as it enables the exploration of novel coupled multipolar orders. Such materials can exhibit distinct phase transitions, topological excitations, and nonlinear responses, potentially leading to new functionalities in multiferroics and magnetoelectric applications.} Specifically, we presented a scenario for cross-coupling between magnetic and electric quadrupole moments, as well as magnetic octupole moments, leading to ferroelectricity. This proposed mechanism can be readily extended to other spin-orbit coupled systems, \SB{including heavy-fermionic compound such as URu$_2$Si$_2$, and might be relevant to the puzzle of HO phase transitions}. Our results lead to a refined classification of multiferroics, incorporating the multipolar sector, and opening new avenues for exploring the coupling of multipole moments beyond those listed in Table~\ref{tab:multi_ferroic}. 

\textit{Note added} \textbf{\---} A notable example is quadrupole-induced ferroelectricity, as discussed in a recent preprint~\cite{Zhao2024}, which aligns with and exemplifies the functionality we describe as ``multipolar multiferroicity."

\textit{Acknowledgments.}\textbf{\---} We acknowledge helpful discussions with \SB{Gabriel Aeppli, Sang-Wook Cheong, Piers Coleman, Niels Grønbech-Jensen, Hosho Katsura, Giniyat Khaliullin, Naoto Nagaosa, and Wei Zhu}. S.B. is grateful to the Department of Physics, University of Connecticut, Storrs, USA, for hosting, where part of the work was initiated. The work in Los Alamos was carried out under the auspices of the U.S. Department of Energy (DOE), Office of Science, and Office of Advanced Scientific Computing Research through the Quantum Internet to Accelerate Scientific Discovery Program. S.B. gratefully acknowledges partial support from the Office of Basic Energy Sciences, Material Sciences and Engineering Division, U.S. DOE, under Contract No. DE-FG02-99ER45790. Work of A.V.B was supported by U.S. Department of Energy, Office of Science, Office of Basic Energy Sciences under award number DE-SC- 0025580 and by European Research Council under the European Union Seventh Framework ERS-2018-SYG 810451 HERO. Computations were performed \SB{(S.H.)} on the cluster of the Center for Materials Theory at the Department of Physics and Astronomy at Rutgers University.

%---------------------------
\bibliographystyle{apsrev4-2}
\bibliography{References}
%---------------------------

%%%%%%%%%%%%%%%% SUPPLEMENTARY %%%%%%%%%%%%%%%%

\clearpage
\pagebreak
\onecolumngrid

\begin{center}
\textbf{\large Supplementary material \--- Multipolar multiferroics in $4d^2/5d^2$ Mott insulators}\\[.2cm]
Saikat Banerjee,$^{1,2,*}$ Stephan Humeniuk,$^{1,\dagger}$ Alan R. Bishop,$^2$ Avadh Saxena,$^2$ and Alexander V. Balatsky$^{3,4}$ \\[.1cm]
{\itshape 
${}^1$ Center for Materials Theory, Rutgers University, Piscataway, New Jersey, 08854, USA \\
${}^2$ Theoretical Division, Los Alamos National Laboratory, Los Alamos, New Mexico 87545, USA \\
${}^3$ Nordita, KTH Royal Institute of Technology and Stockholm University, 106 91 Stockholm, Sweden \\
${}^4$ Department of Physics, University of Connecticut, Storrs, Connecticut 06269, USA \\}
(Dated: \today)\\[1cm]
\end{center}

\setcounter{equation}{0}
\setcounter{figure}{0}
\setcounter{table}{0}
\setcounter{page}{1}
\renewcommand{\theequation}{S\arabic{equation}}
\renewcommand{\thefigure}{S\arabic{figure}}
\renewcommand{\bibnumfmt}[1]{[S#1]}
\renewcommand{\citenumfont}[1]{S#1}

\tableofcontents 

%----------------------------------------------------------------------
\section{Introduction: Hubbard-Kanamori description \label{sec:sec_1}}
%----------------------------------------------------------------------

In this section, we delve into the specifics of the microscopic model Hamiltonian governing the behavior of transition metal (TM) compounds characterized by $d^2$-electronic configurations. The comprehensive Hamiltonian indicated in the main text is expressed as follows 
%---------------------------
\begin{subequations}
\begin{align}
\label{seq.1.1}
\mathcal{H} 	& = \mathcal{H}_{0} + \mathcal{H}_{1} + \mathcal{H}_{\rm{crys.}}, \\
\label{seq.1.2}
\mathcal{H}_0 	& = U\sum_{i\alpha} n_{i\alpha,\uparrow}n_{i\alpha,\downarrow} + \frac{U'}{2}\sum_{\substack{i,\alpha\neq \beta, \\ \sigma, \sigma'}} n_{i\alpha \sigma} n_{i\beta \sigma'}  -\frac{J_\mathrm{H}}{2}\sum_{\substack{\alpha\neq \beta, \\ \sigma, \sigma'}} d^{\dagger}_{i\alpha \sigma} d_{i\alpha\sigma'} d^{\dagger}_{i\beta \sigma'}  d_{i\beta \sigma}^{\phantom\dagger} + \lambda \sum_{i} d^{\dagger}_{i\alpha \sigma} \left( \mathbf{L}_{\alpha \beta} \cdot \mathbf{S}_{\sigma \sigma'} \right) d_{i \beta \sigma'}, \\
\label{seq.1.3}
\mathcal{H}_1 	& = \sum_{\substack{\langle ij \rangle \\ \alpha \beta \sigma \sigma'}} t_{\alpha\beta; \sigma \sigma'} d^{\dag}_{i \alpha \sigma} d_{j \beta \sigma'} + 
t_{pd} \sum_{\substack{i,l \\ \alpha \sigma}} p^{\dag}_{l \sigma} d_{i\alpha \sigma} +  \mathrm{h.c.},
\end{align}
\end{subequations}
%---------------------------
where (i) the indices $\alpha$ and $\beta$ represent orbital indices, and they can take on values from the set $\{xy, yz, zx\}$, (ii) the $\sigma$ term corresponds to the electronic spin degrees of freedom, (iii) $U$ represents the strength of the onsite Coulomb repulsion, and (iv) $J_{\rm{H}}$ stands for Hund's coupling. In rotationally invariant systems, $U' = U - 2 J_{\rm{H}}$, (iv) $\lambda$ denotes the strength of the atomic spin-orbit coupling (SOC), and (v) the operator $p^{\dag}_\sigma$ creates an electron at the ligand site with spin $\sigma$. In addition, a crystal field term $\mathcal{H}_{\rm{crys.}}$, which accounts for the charge-transfer energy gap between the $d$- and $p$-sites, is introduced to the system as $\mathcal{H}_{\rm{crys.}}=\Delta_{\rm{crys.}} \sum_{l} p^{\dag}_{l\sigma} p_{l \sigma}$. Here, $t_{\alpha \beta; \sigma \sigma'}$ characterizes the hopping amplitude between two TM sites, while $t_{pd}$ represents the hopping amplitude between a TM site and a ligand site.

%-------------------------------------------------------------------------------------------
\begin{figure}[b!]
\centering \includegraphics[width=0.6\linewidth]{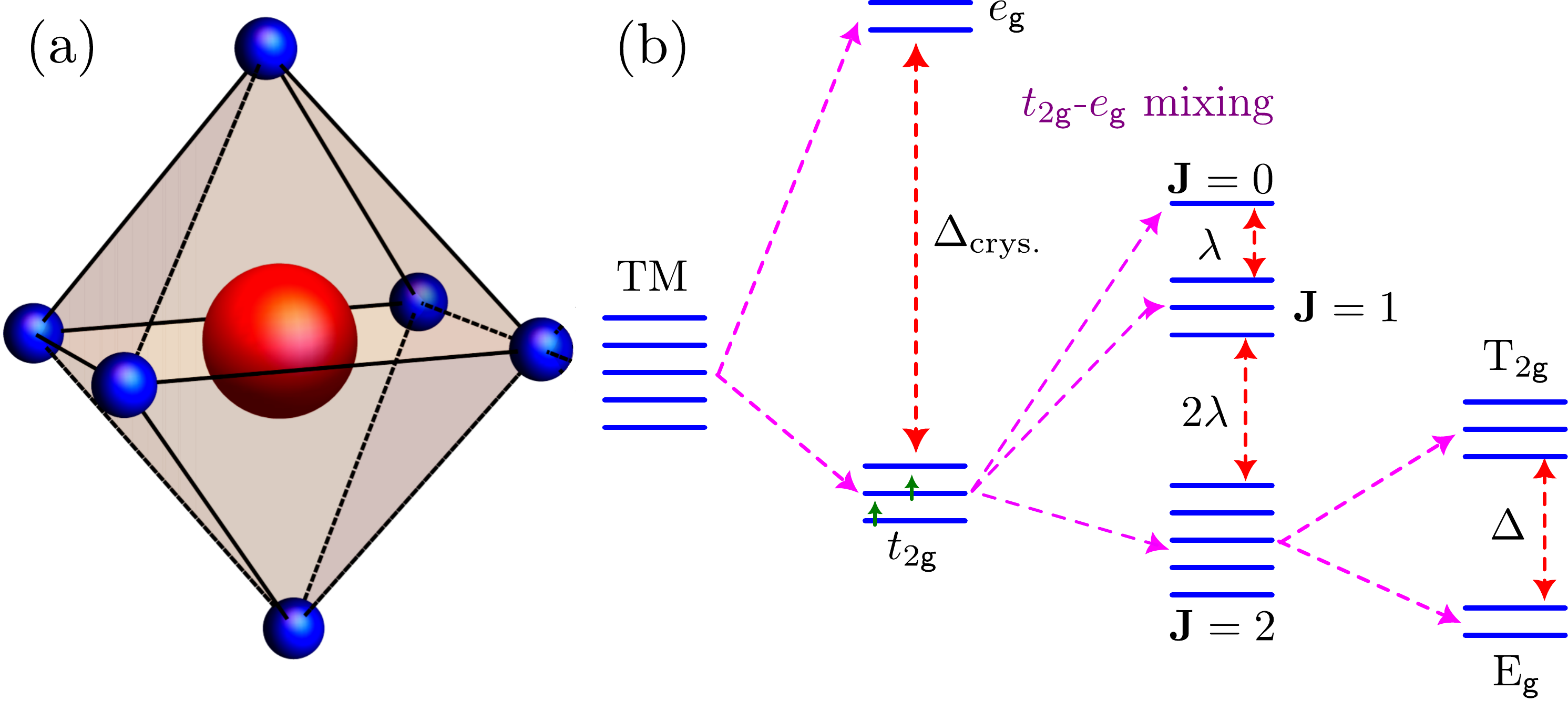}
\caption{(a) Schematic of the octahedral unit cell containing a central transition metal (TM) ion surrounded by six ligand sites. (b) Hierarchy of atomic onsite energy splitting for the five degenerate $d$-levels resulting from various interactions, as discussed in the text.}\label{fig:SFig1}
\end{figure}
%-------------------------------------------------------------------------------------------

Before moving forward, we first analyze the onsite Hamiltonian $\mathcal{H}_0$ in detail to characterize the overall energy splitting. For this purpose, we strictly follow the following energy hierarchy as $U \gg \Delta_{\rm{crys.}} \gg J_{\rm{H}} \gtrsim \lambda$. Since there are two electrons in the $d^2$-configuration, the total spin of the two-electron state is $\bf{S} = 1$. We consider a large crystal field in octahedral geometry, as shown in Fig. 1 in the main text. This leads to the conventional $t_{2\dg}$ and $e_{\dg}$ splitting of the isolated five degenerate $d$ levels as shown in Fig.~\ref{fig:SFig1}(b). However, because of the SOC the low-energy $t_{2\dg}$ manifold further splits into total angular momentum ${\bf{J}} = ({\bf{L}} + {\bf{S}}) = 0,1,2$ sectors. Since the ${\bf{J}} = 2$ sector transforms under the same symmetry as the original five $d$-levels, it can be shown that the effective renormalized crystal field further splits this low-energy ${\bf{J}} = 2$ manifold into yet another splitting. Considering the same terminology as in the previous case, we associate this to ${\rm{T}}_{2\dg}$ and ${\rm{E}}_{\dg}$ levels [see Fig.~\ref{fig:SFig1}(b) for an illustration]~\cite{Takayama2021}. Hence, the two-electron ground state is composed of a non-Kramers doublet labeled by ${\rm{E}}_{\dg}$. 

In an octahedral environment embedded in cubic geometry, we can formally write down the effective residual crystal field Hamiltonian as 
%-----------------------------
\begin{equation}\label{seq.2}
\mathcal{H}_{\rm{res. crys. fld.}} = - \Delta \left( \mathcal{O}_{40} + 5 \mathcal{O}_{44} \right),
\end{equation}
%-----------------------------
where $\mathcal{O}_{mn}$ are the Stevens operators~\cite{PhysRevB.101.054439,PhysRevB.101.155118} written in the $J$-basis as
%--------------------
\begin{subequations}
\begin{align}
\label{seq.3.1}
\mathcal{O}_{40} & = 35 J_z^4 - [30 J (J+1) - 25]J_z^2 + 3J^2(J+1)^2 - 6J (J+1), \\
\label{seq.3.2}
\mathcal{O}_{44} & = \frac{1}{2} \left( J_+^4 + J_-^4 \right),
\end{align}
\end{subequations}
%--------------------
where $J_{\pm} = J_x \pm i J_y$ are the raising and lowering operators, respectively, and $\Delta$ is the renormalized strength of the SOC-enabled crystal fields. Note that the octahedral group is a subgroup of the cubic group. Before moving forward, we first analyze the energy manifold constructed out of the two-electron states in the $d^2$-Mott insulators. 

%--------------------------------------------------------------
\subsection{Energy manifold with strong SOC \label{sec:sec.1.1}}
%--------------------------------------------------------------

We consider a large crystal field compared to the parameters in the Hamiltonian in Eq.~\eqref{seq.1.2}, subsequently leading to a large $t_{2\dg}$ and $e_\dg$ splitting in the $d$-orbital manifold. Therefore, in the $d^2$-configuration, two electrons are forced to stay in the $t_{2\dg}$ manifold, which is composed of three orbital angular momentum active orbitals: $d_{xy}$, $d_{yz}$, and $d_{zx}$. Since they are pointed orbitals and also form a cubic coordinate system, following Ref.~\cite{PhysRevLett.111.197201}, we relabel them as $a \rightarrow d_{yz}$, $b \rightarrow d_{zx}$, and $c \rightarrow d_{xy}$. The two electrons reside in the $t_{2g}$ manifold without any SOC. Therefore, there are three different configurations. Here, we consider $\Delta_{\rm{crys}} > U \gg J_{\rm{H}}, \lambda$. Hence, following the Hund's rule, we have
%----------------------------
\begin{equation}\label{seq.4}
\mathrm{A} = \{ b c \}, \;  \mathrm{B} = \{ a c \}, \;  \mathrm{C} = \{ a b \}, \; 
\end{equation}
%----------------------------
where we have ignored the spin degrees of freedom. As labeled above, the composite orbital configurations also form a cubic coordinate system. Writing out explicitly, we have
%----------------------------
\begin{equation}\label{seq.5}
\ket{\psi_{i;\alpha \beta, \sigma \sigma'}} = d^{\dagger}_{i;\alpha \sigma} d^{\dagger}_{i;\beta \sigma'} \ket{0}, \quad \forall \; \alpha, \beta \in \{xy, yz, zx \},
\end{equation}
%----------------------------
where $d^{\dagger}_{i; \alpha, \sigma}$ creates one electron at the site $i$ with orbital $\alpha$, and spin $\sigma$, and $\ket{0}$ corresponds to the empty $t_{2\dg}$ manifold. There are $^6C_2$ = $15$ possible configurations following Eq.~\eqref{seq.5}. To compute the energy, we use the simplified form for the Hubbard-Kanamori model (HKM) for the rotation invariant $t_{2\dg}$ manifold as~\cite{PhysRevB.95.014409,Georges2013}
%----------------------------
\begin{equation}\label{seq.6}
\mathcal{H}^{{(i)}}_{0} = (U-3J_{\rm{H}}) \frac{N_i(N_i - 1)}{2} + J_{\rm{H}} \left(\frac{5}{2}N_i - 2S_i^2 -\frac{1}{2}L_i^2 \right),
\end{equation}
%----------------------------
where $N_i$ is electron number, $S_i$ is total spin, and $L_i$ is total orbital angular momentum. We include the SOC $\lambda$ into the HKM. In this case, the spin and orbital angular momenta are entangled, and one needs to work in the total angular momentum basis as both $\mathbf{L}_i$ and $\mathbf{S}_i$ are not good quantum numbers anymore. The SOC part can be written as~\cite{PhysRevLett.110.087203}
%-----------------------------
\begin{equation}\label{seq.7}
\mathcal{H}^{(i)}_{\rm{soc}} = \lambda \mathbf{L}_i \cdot \mathbf{S}_i = \frac{\lambda}{2} \left( \mathbf{J}_i^2 - \mathbf{L}_i^2 -\mathbf{S}_i^2 \right),
\end{equation}
%-----------------------------
where $\mathbf{J}_i = \mathbf{L}_i + \mathbf{S}_i$, with $\mathbf{L}_i$, and $\mathbf{S}_i$ being the total orbital and spin angular momentum, respectively. In this case, the previous six-fold degeneracy is broken. We have a further splitting of $t_{2\dg}$ manifold in $J = 0,1,2$ sub-manifolds, with the respective energies dictated by $\mathcal{H}^{(i)}_{\rm{soc}}$ as: $E^{(i)}_{\rm{soc}}|_{J=0} = -2\lambda$, $E^{(i)}_{\rm{soc}}|_{J=1} = -\lambda$, and $E^{(i)}_{\rm{soc}}|_{J=2} = \lambda$, where $J = 0$, $J = 1$, and $J = 2$ are singly, triply, and quintuply degenerate manifolds, respectively. 

%-----------------------------------------------------------------------
\subsection{Representation of the low-energy states \label{sec:sec.1.2}}
%-----------------------------------------------------------------------

At first, we define the orbital angular momentum operators on the basis of A, B, and C defined in Eq.~\eqref{seq.6} as follows~\cite{PhysRevLett.111.197201}
%----------------------------
\begin{equation}\label{seq.8}
L^{(i)}_x = -i \left( \rm{B}^{\dag} \rm{C} - \rm{C}^{\dag} \rm{B} \right), \quad
L^{(i)}_y = -i \left( \rm{C}^{\dag} \rm{A} - \rm{A}^{\dag} \rm{C} \right), \quad
L^{(i)}_z = -i \left( \rm{A}^{\dag} \rm{B} - \rm{B}^{\dag} \rm{A} \right).
\end{equation}
%----------------------------
For later convenience, it is better to introduce the raising, lowering, and orbital angular momentum preserving operators acting on the empty $t_{2\dg}$ manifold as~\cite{Kumar2022}
%--------------------
\begin{equation}\label{seq.9}
\bd^{\dagger}_{i;0} \ket{0} = {\rm{C}}^{\dagger}_{i} \ket{0}, \quad
\bd^{\dagger}_{i;+} \ket{0} = \frac{1}{\sqrt{2}} \left( {\rm{A}}^{\dagger}_{i} + i {\rm{B}}^{\dagger}_{i} \right) \ket{0}, \quad
\bd^{\dagger}_{i;-} \ket{0} = \frac{1}{\sqrt{2}} \left( {\rm{A}}^{\dagger}_{i} - i {\rm{B}}^{\dagger}_{i} \right) \ket{0}, 
\end{equation}
%--------------------
where we consider the implicit spin dependence on the operators defined above in Eq.~\eqref{seq.7}. It is straightforward to check that the states defined above have fixed orbital angular momentum as 
%--------------------
\begin{equation}\label{seq.10}
\bra{0}\bd_{i;0}| L^{(i)}_z | \bd^{\dagger}_{i;0} \ket{0}  = 0, \quad
\bra{0} \bd_{i;+}| L^{(i)}_z | \bd^{\dagger}_{i;+} \ket{0} = 1, \quad
\bra{0} \bd_{i;-}| L^{(i)}_z | \bd^{\dagger}_{i;-} \ket{0} = -1.
\end{equation}
%--------------------
Finally, we focus on the effective spin angular momentum states. The effective total spin is $\mathbf{S}_{\rm{eff}} = 1$ with the three spin components as $\ket{m_s = 1}$, $\ket{m_s = -1}$, and $m_s = 0$. The first two states are easily written in the $\bd$ operator basis 
%--------------------
\begin{equation}\label{seq.11}
\bd^{\dagger}_{i;\gamma;\uparrow \uparrow} \ket{0}, \; \forall m_s = 1, \;
\bd^{\dagger}_{i;\gamma;\downarrow \downarrow} \ket{0}, \; \forall m_s = -1 \quad (\gamma = \{0, \pm 1 \}),
\end{equation}
%--------------------
where $\uparrow,\downarrow$ corresponds to the specific spin-orientation in Eq.~\eqref{seq.7}, such that for a state with $m_l = 1$, and $m_s  = 1$, we can write it as
%-----------------------------
\begin{equation}\label{seq.12}
\bd^{\dagger}_{i;+,\uparrow \uparrow} 
= \frac{1}{\sqrt{2}} 
\left( d^{\dagger}_{i;xy \uparrow} d^{\dagger}_{i;zx \uparrow}
+ 
i d^{\dagger}_{i;xy \uparrow} d^{\dagger}_{i;yz \uparrow} \right) \ket{0}.
\end{equation}
%-----------------------------
The other states can be written in a similar fashion. For latter convenience, we rewrite all the low-lying nine states [see Fig.~\ref{fig:SFig1}(b)] in the $\ket{m_l,m_s}$ basis as
%---------------------
\begin{subequations}
\begin{align}
\label{seq.13.1}
\ket{1,1} & = 
\frac{1}{\sqrt{2}} 
\left( d^{\dagger}_{i;xy \uparrow} d^{\dagger}_{i;zx \uparrow} + i d^{\dagger}_{i;xy \uparrow} d^{\dagger}_{i;yz \uparrow} \right) \ket{0}, \\
\label{seq.13.2}
\ket{1,0} & = 
\frac{1}{2} 
\left( d^{\dagger}_{i;xy \uparrow} d^{\dagger}_{i;zx \downarrow} + d^{\dagger}_{i;xy \downarrow} d^{\dagger}_{i;zx \uparrow} + i d^{\dagger}_{i;xy \uparrow} d^{\dagger}_{i;yz \downarrow} + i d^{\dagger}_{i;xy \downarrow} d^{\dagger}_{i;yz \uparrow}\right) \ket{0}, \\
\label{seq.13.3}
\ket{1,-1} & = 
\frac{1}{\sqrt{2}} \left( d^{\dagger}_{i;xy \downarrow} d^{\dagger}_{i;zx \downarrow} + i d^{\dagger}_{i;xy \downarrow} d^{\dagger}_{i;yz \downarrow} \right) \ket{0}, \\
\label{seq.13.4}
\ket{-1,1} & = 
\frac{1}{\sqrt{2}} \left( d^{\dagger}_{i;xy \uparrow} d^{\dagger}_{i;zx \uparrow} - i d^{\dagger}_{i;xy \uparrow} d^{\dagger}_{i;yz \uparrow} \right) \ket{0}, \\
\label{seq.13.5}
\ket{-1,0} & = 
\frac{1}{2} \left( d^{\dagger}_{i;xy \uparrow} d^{\dagger}_{i;zx \downarrow} + d^{\dagger}_{i;xy \downarrow} d^{\dagger}_{i;zx \uparrow} - i d^{\dagger}_{i;xy \uparrow} d^{\dagger}_{i;yz \downarrow} - i d^{\dagger}_{i;xy \downarrow} d^{\dagger}_{i;yz \uparrow}\right) \ket{0}, \\
\label{seq.13.6}
\ket{-1,-1} & = 
\frac{1}{\sqrt{2}} \left( d^{\dagger}_{i;xy \downarrow} d^{\dagger}_{i;zx \downarrow} - i d^{\dagger}_{i;xy \downarrow} d^{\dagger}_{i;yz \downarrow} \right) \ket{0}, \\
\label{seq.13.7}
\ket{0,1} & = 
d^{\dag}_{i;yz, \uparrow} d^{\dagger}_{i; zx, \uparrow} \ket{0}, \\
\label{seq.13.8}
\ket{0,0} & = 
\frac{1}{\sqrt{2}} \left( d^{\dagger}_{i;yz \uparrow} d^{\dagger}_{i;zx \downarrow} +  d^{\dagger}_{i;yz \downarrow} d^{\dagger}_{i;zx \uparrow} \right) \ket{0}, \\
\label{seq.13.9}
\ket{0,-1} & = 
d^{\dag}_{i;yz, \downarrow} d^{\dagger}_{i; zx, \downarrow} \ket{0}.
\end{align}
\end{subequations}
%---------------------
Note that the SOC is comparable to and often exceeds the strength of the Hund's coupling. This subsequently leads to the three different states in the Hund's + SOC submanifold [Eq.~\eqref{seq.13.1}-Eq.~\eqref{seq.13.9}] corresponding to ${\bf{J}} = 2$ (pseudo-spin-quintuplet), ${\bf{J}} = 1$ (pseudo-spin-triplet), and ${\bf{J}} = 0$ (pseudo-spin-singlet) sectors~\cite{Takayama2021}. Finally, the ${\bf{J}} = 2$ states further split into the T$_{2\dg}$, and E$_\dg$ manifold as explained earlier in Eq.~\eqref{seq.2}. 

%-------------------------------------------------------------------------------
\subsection{Clebsch-Gordon coefficients: Individual states \label{sec:sec.1.3}}
%-------------------------------------------------------------------------------

Now, we rewrite the atomic singlet, triplet, and quintuplet states (in the $\bf{J}$ basis) in terms of the $\ket{m_l,m_s}$ basis utilizing the Clebsch-Gordon coefficients. Consequently, we obtain 
%---------------------
\begin{subequations}
\begin{align}
\label{seq.14.1}
\ket{\mathbf{J}_i = 0; J_{i;z} = 0}	
& = \frac{1}{\sqrt{3}} \left( \ket{1,-1} - \ket{0,0} + \ket{-1,1}\right), \\
\label{seq.14.2}
\ket{\mathbf{J}_i = 1; J_{i;z} = +1}	
& = \frac{1}{\sqrt{2}} \left( \ket{1,0} - \ket{0,1} \right), \\
\label{seq.14.3}
\ket{\mathbf{J}_i = 1; J_{i;z} = 0} 
& = \frac{1}{\sqrt{2}} \left( \ket{1,-1} - \ket{-1,1} \right), \\
\label{seq.14.4}
\ket{\mathbf{J}_i = 1; J_{i;z} = -1}	
& = -\frac{1}{\sqrt{2}} \left( \ket{-1,0} - \ket{0,-1} \right), \\
\label{seq.14.5}
\ket{\mathbf{J}_i = 2; J_{i;z} = +2}	
& = \ket{1,1}, \\
\label{seq.14.6}
\ket{\mathbf{J}_i = 2; J_{i;z} = +1}	
& = \frac{1}{\sqrt{2}} \left( \ket{1,0} + \ket{0,1} \right), \\
\label{seq.14.7}
\ket{\mathbf{J}_i = 2; J_{i;z} = 0}	
& = \frac{1}{\sqrt{6}} \left( \ket{1,-1} + 2 \ket{0,0} + \ket{-1,1} \right), \\
\label{seq.14.8}
\ket{\mathbf{J}_i = 2; J_{i;z} = -1}	
& = -\frac{1}{\sqrt{2}} \left( \ket{-1,0} + \ket{0,-1} \right), \\
\label{seq.14.9}
\ket{\mathbf{J}_i = 2; J_{i;z} = -2}	
& = \ket{-1,-1}.
\end{align}
\end{subequations}
%---------------------
Focusing on the energy splitting as shown in Fig.~\ref{fig:SFig1}(b) and discussed earlier, the lowest doublet ${\rm{E}}_{\dg}$ can be written explicitly with the above states as
%-----------------------------
\begin{equation}\label{seq.15}
\ket{{\rm{E}}_{\dg}} : \quad \ket{\Uparrow} = \frac{\ket{\mathbf{J}_i = 2; J_{i;z} = +2} + \ket{\mathbf{J}_i = 2; J_{i;z} = -2}}{\sqrt{2}}, \quad \ket{\Downarrow} = \ket{\mathbf{J}_i = 2; J_{i;z} = 0}.
\end{equation}
%-----------------------------
Using the two-electron states in Eq.~\eqref{seq.13.1}, Eq.~\eqref{seq.13.6}, and Eq.~\eqref{seq.13.8} in Eq.~\eqref{seq.15}, one can obtain the doublet states as in Eq.~(1) of the main text. Note that the above two pseudo-spin states are non-Kramers doublet. In terms of the original spin-orbit coupled $d$-electron operators, the above two pseudo-spin states are written as 
%---------------------
\begin{subequations}
\begin{align}
\label{seq.16.1}
\ket{i;\Uparrow} & = 
\frac{1}{2}\left( 
d^{\dag}_{i;xy\uparrow} d^{\dag}_{i;zx\uparrow} + d^{\dag}_{i;xy\downarrow} d^{\dag}_{i;zx\downarrow} + 
i d^{\dag}_{i;xy\uparrow} d^{\dag}_{i;yz\uparrow} - i d^{\dag}_{i;xy\downarrow} d^{\dag}_{i;yz\downarrow}
\right) \ket{0}, \\
\label{seq.16.2}
\ket{i;\Downarrow} & = 
\frac{1}{2\sqrt{3}}\left(
d^{\dag}_{i;xy\downarrow} d^{\dag}_{i;zx\downarrow} + d^{\dag}_{i;xy\uparrow} d^{\dag}_{i;zx\uparrow} + 
2d^{\dag}_{i;yz\uparrow} d^{\dag}_{i;zx\downarrow} + 2d^{\dag}_{i;yz\downarrow} d^{\dag}_{i;zx\uparrow} + 
i d^{\dag}_{i;xy\downarrow} d^{\dag}_{i;yz\downarrow} - i d^{\dag}_{i;xy\uparrow} d^{\dag}_{i;yz\uparrow} 
\right) \ket{0}.
\end{align}
\end{subequations}
%---------------------
%--------------------------------------------------------------------------------------------
\section{Schrieffer-Wolff transformation: Multipolar exchange Hamiltonian} \label{sec:sec.2}
%--------------------------------------------------------------------------------------------

In the previous section, we described the origin of low-energy effective pseudo-spin [non-Kramers doublet, ${\rm{E}}_{\dg}$] in the atomic limit of each TM atom. However, there will always be virtual electron transfer between the neighboring TM sites in a TM compound. Such virtual exchanges lead to an effective interaction between the emergent pseudo-spin states. Before addressing the emergence of such an exchange interaction, we first notice that the two non-Kramers states in Eq.~\eqref{seq.15} host the non-vanishing expectation value of certain higher-order (multipolar) operators. It is easy to see that only three Stevens operators can have non-vanishing expectation values within these low-energy states. We define these higher-order (multipolar) moments as~\cite{Hoffmann_1991}
%-----------------------------
\begin{equation}\label{seq.17}
\mathcal{O}_{20} = 3J_z^2 - {\bf{J}}^2,  \quad
\mathcal{O}_{22} = J_x^2 - J_y^2, \quad 
\mathsf{T}_{xyz} = \overline{J_x J_y J_z}.
\end{equation}
%-----------------------------
Note that the first two operators are quadrupolar, and the last one is an octupolar moment (the overline signifies the full symmetrization over the directions $\{x,y,z\}$). The low-energy ground states are manifestly dipole inactive. It is easy to check that the above operators have finite expectation values between the pseudo-spin up and down states as
%-----------------------------
\begin{subequations}
\begin{align}
\label{seq.18.1}
&	\braket{i;\Uparrow|\mathcal{O}_{20}|i;\Uparrow} = 6, 						& 	\braket{i;\Downarrow|\mathcal{O}_{20}|i;\Downarrow} &	= -6, \\
\label{seq.18.2}
&	\braket{i;\Uparrow|\mathcal{O}_{22}|i;\Downarrow} = 2\sqrt{3}, 				& 	\braket{i;\Downarrow|\mathcal{O}_{22}|i;\Uparrow} 	&	= 2\sqrt{3}, \\
\label{seq.18.3}
&	\braket{i;\Uparrow|\mathsf{T}_{xyz}|i;\Downarrow}  =  -i \sqrt{3}, 		&	\braket{i;\Downarrow|\mathsf{T}_{xyz}|i;\Uparrow} &	= i \sqrt{3}.
\end{align}
\end{subequations}
%-----------------------------
Upon suitable renormalization, we map the three Stevens operators to SU(2) generators by considering the standard Pauli matrices as $(\tilde{\sigma}^x, \tilde{\sigma}^y, \tilde{\sigma}^z)$. The normalization becomes evident from the above relations as
%-----------------------------
\begin{equation}\label{seq.19}
\frac{\mathcal{O}_{22}}{4\sqrt{3}} \rightarrow \tilde{\sigma}^x, 	\qquad
\frac{\mathsf{T}_{xyz}}{2\sqrt{3}} \rightarrow \tilde{\sigma}^y, 	\qquad
\frac{\mathcal{O}_{20}}{12} \rightarrow \tilde{\sigma}^z.
\end{equation}
%-----------------------------
However, these pseudo-spin generators are distinct from the conventional spin-$\tfrac{1}{2}$ representation. Most importantly, the transformation property under time-reversal symmetry (TRS) transformation ($\mathcal{T}$) is distinct, as is evident from the form of the multipolar moments, and we obtain 
%-----------------------------
\begin{equation}\label{seq.20}
{\cal{T}}: \; 
\{ \tilde{\sigma}^x, \tilde{\sigma}^y, \tilde{\sigma}^z \} 
\rightarrow 
\{ \tilde{\sigma}^x, - \tilde{\sigma}^y, \tilde{\sigma}^z \}.
\end{equation}
%-----------------------------

%--------------------------------------------------------------------------------------------
\subsection{Derivation of the low-energy effective exchange interaction \label{sec:sec.2.1}}
%--------------------------------------------------------------------------------------------

Up to this point, we have described the low-energy behavior of the multipolar degrees of freedom at each site. However, due to the virtual exchange of electrons, these individual multipolar moments develop couplings with neighboring sites. The Schrieffer-Wolff transformation (SWT) is a powerful tool that systematically derives a low-energy exchange Hamiltonian that captures these couplings. Interested readers can refer to Refs.~\cite{PhysRev.149.491,PhysRev.157.295} for a more detailed explanation of SWT. However, for a self-contained analysis, we will concisely describe SWT. We begin with the total Hamiltonian $\mathcal{H} = \mathcal{H}_0 + \mathcal{H}_1$, where $\mathcal{H}_0$ typically represents the correlated part, and $\mathcal{H}_1$ is considered a small perturbation. We then perform a unitary transformation as follows
%-------------------------
\begin{align}\label{seq.21}
\mathcal{H}' & = U^{\dagger} \mathcal{H} U = e^{i\bs} \mathcal{H}e^{-i\bs}, \\
& = \mathcal{H} + \big[i\bs, \mathcal{H} \big] + \frac{1}{2!} \big[i\bs, \big[ i\bs,\mathcal{H} \big]\big] + \frac{1}{3!} \big[i\bs, \big[i\bs, \big[i\bs, \mathcal{H} \big] \big] \big] + \cdots, \nonumber \\
& = \mathcal{H}_0 + \mathcal{H}_1 + i\big[\bs, \mathcal{H}_0] + i\big[\bs, \mathcal{H}_1] - \frac{1}{2}\big[\bs, \big[ \bs,\mathcal{H}_0 \big]\big] - \frac{1}{2} \big[\bs, \big[\bs,\mathcal{H}_1 \big]\big] - \frac{i}{3!} \big[\bs, \big[\bs, \big[\bs, \mathcal{H}_0 \big] \big] \big] - \frac{i}{3!} \big[\bs, \big[\bs, \big[\bs, \mathcal{H}_1 \big] \big] \big] + \cdots, \nonumber
\end{align}
%-------------------------
where $\bs$ is a hermitian operator. We now write the generator $\bs$ in a perturbative expansion with the hopping parameters as $\bs = \bs^{(1)} + \bs^{(2)} + \bs^{(3)} + \bs^{(4)} + \cdots $. Now, plugging this back in Eq.~\eqref{seq.20} and collecting terms of the same order in expansion, we obtain 
%-------------------------
\begin{align}\label{seq.22}
\mathcal{H}' & = \mathcal{H}_0 + \mathcal{H}_1 + i \big[ \bs^{(1)}, \mathcal{H}_0 \big] \nonumber \\
& + i \big[ \bs^{(1)}, \mathcal{H}_1 \big] + i\big[ \bs^{(2)}, \mathcal{H}_0 \big] - \frac{1}{2} \big[\bs^{(1)}, \big[ \bs^{(1)}(t), \mathcal{H}_0 \big]\big] \nonumber \\
& + i \big[ \bs^{(3)}, \mathcal{H}_0 \big] + i\big[ \bs^{(2)}, \mathcal{H}_1 \big] - \frac{1}{2} \big[ \bs^{(1)}, \big[ \bs^{(1)}, \mathcal{H}_1 \big] + \big[ \bs^{(2)} , \mathcal{H}_0 \big]\big] - \frac{1}{2} \big[\bs^{(2)}, \big[ \bs^{(1)}, \mathcal{H}_0 \big]\big] - \frac{i}{3!} \big[ \bs^{(1)}, \big[ \bs^{(1)}, \big[ \bs^{(1)}, \mathcal{H}_0 \big]\big]\big] \nonumber \\
& + i \big[ \bs^{(4)}, \mathcal{H}_0 \big] + i\big[ \bs^{(3)}, \mathcal{H}_1 \big] - \frac{1}{2} \big[ \bs^{(3)}, \big[ \bs^{(1)}, \mathcal{H}_0\big] \big] -\frac{1}{2} \big[ \bs^{(2)}, \big[ \bs^{(1)}, \mathcal{H}_1 \big] + \big[ \bs^{(2)}, \mathcal{H}_0 \big]\big]  -\frac{1}{2} \big[\bs^{(1)}, \big[ \bs^{(2)}, \mathcal{H}_1 \big] + \big[ \bs^{(3)}, \mathcal{H}_0 \big] \big] \nonumber \\
& -\frac{i}{3!} \big[ \bs^{(1)}, \big[ \bs^{(1)}, \big[ \bs^{(1)}, \mathcal{H}_1 \big] \big]  + \big[ \bs^{(1)}, \big[ \bs^{(2)}, \mathcal{H}_0\big] \big] + \big[ \bs^{(2)}, \big[ \bs^{(1)}, \mathcal{H}_0\big] \big]\big] - \frac{i}{3!}  \big[ \bs^{(2)}, \big[ \bs^{(1)}, \big[ \bs^{(1)}, \mathcal{H}_0\big] \big] \nonumber \\
&  + \frac{1}{4!}  \big[ \bs^{(1)}, \big[ \bs^{(1)}, \big[ \bs^{(1)}, \big[ \bs^{(1)}, \mathcal{H}_0 \big]\big]\big] \big] + \mathcal{O}(S^{(5)}) \ldots \,. 
\end{align}
%-----------------------------
The evaluation of the effective Hamiltonian at each perturbative order becomes evident from Eq.~\eqref{seq.21}. We focus on the terms up to the $n$-th order and obtain the low-energy effective Hamiltonians up to the $(n+1)$-th order, along with the corresponding generating functions $\bs^{(n)}$. Here, we restrict ourselves to the generic expressions for the first two lowest-order effective Hamiltonians below. These will be utilized later to derive the multipolar exchange Hamiltonians. However, for the explicit evaluation of the generating functions $\bs^{(n)}$ at each order, we need to introduce the projection operators $\mathcal{P}$ and $\mathcal{Q}$. These operators act on the low- and high-energy sectors of the total Hilbert space, respectively. \\

\textbf{Second-order effective Hamiltonian \---} The Hamiltonian can be determined up to the second order in perturbation parameters by utilizing Eq.~\eqref{seq.21} (assuming $\bs^{(2)} = 0$), and it is given by
%-----------------------------
\begin{equation}\label{seq.23}
\mathcal{H}^{(2)}_{\mathrm{eff}} = \frac{i}{2} \big[ \bs^{(1)}, \mathcal{H}_1\big].
\end{equation}
%-----------------------------
The dynamic equation governing the generating function $\bs^{(1)}$ can be derived from the first-order terms in Eq.~\eqref{seq.21}, leading to
%-----------------------------
\begin{equation}\label{seq.24}
\big[ \bs^{(1)}, \mathcal{H}_0 \big]  = i\mathcal{H}_1.
\end{equation}
%-----------------------------
The equation presented above is a matrix equation, where $\bs^{(1)}$ is a $2 \times 2$ matrix defined within the space of projectors $\mathcal{P}$ and $\mathcal{Q}$, and can be expressed as
%-----------------------------
\begin{equation}\label{seq.25}
\bs^{(1)}  = \begin{pmatrix}
\mathcal{P} \bs^{(1)} \mathcal{P} & \mathcal{P} \bs^{(1)} \mathcal{Q} \\
\mathcal{Q} \bs^{(1)} \mathcal{P} & \mathcal{Q} \bs^{(1)} \mathcal{Q}
\end{pmatrix}.
\end{equation}
%-----------------------------
As we are primarily concerned with the low-energy states, the diagonal elements are set to zero, and our focus is solely on solving for the off-diagonal elements. This leads us to the following equation (it is important to note that $\mathcal{H}_0$ is diagonal in the $\mathcal{Q}$ space): 
%--------------------
\begin{subequations}
\begin{align}
& \mathcal{P} \bs^{(1)} \mathcal{H}_0\mathcal{Q} -  \mathcal{P} \mathcal{H}_0 \bs^{(1)} \mathcal{Q} = i\mathcal{P}\mathcal{H}_1 \mathcal{Q} \, \nonumber \\
\label{seq.26.1}
&  \mathcal{P} \bs^{(1)} \mathcal{Q} \mathcal{Q} \mathcal{H}_0 \mathcal{Q} - \cancel{ \mathcal{P} \mathcal{H}_0 \mathcal{Q} \mathcal{Q} \bs^{(1)} \mathcal{Q} } = i \mathcal{P}\mathcal{H}_1 \mathcal{Q}, \, \\
& \mathcal{Q} \bs^{(1)} \mathcal{H}_0\mathcal{P} - \mathcal{Q} \mathcal{H}_0 \bs^{(1)} \mathcal{P} = i \mathcal{Q}\mathcal{H}_1 \mathcal{P} \, \nonumber \\
\label{seq.26.2}
& \cancel{ \mathcal{Q} S^{(1)} \mathcal{Q} \mathcal{Q} \mathcal{H}_0\mathcal{P}} -  \mathcal{Q} \mathcal{H}_0 \mathcal{Q} \mathcal{Q} \bs^{(1)} \mathcal{P} =  i\mathcal{Q}\mathcal{H}_1 \mathcal{P},
\end{align}
\end{subequations}
%--------------------
where $\mathcal{Q} \mathcal{H}_0 \mathcal{Q}$ corresponds to eigenvalues in the intermediate high-energy states. \\

\textbf{Third-order effective Hamiltonian \---} Indeed, let us also outline the steps for obtaining the third-order extension. Starting from Eq. \eqref{seq.21}, we can systematically derive the effective Hamiltonian within the framework of third-order perturbation theory. We examine the algebraic and dynamical equation for the generating function $\bs^{(2)}$ as 
%-----------------------------
\begin{equation}\label{seq.27}
\big[ \bs^{(2)}, \mathcal{H}_0 \big] =  -\big[ \bs^{(1)}, \mathcal{H}_1 \big].
\end{equation}
%-----------------------------
Note that we can simplify the last term in the second-order expansion of Eq. \eqref{seq.21} using the expression for $\bs^{(1)}$ from Eq.~\eqref{seq.23} as follows. Consequently, the effective Hamiltonian up to third-order in perturbation becomes (assuming $\bs^{(3)} = 0$ as before): 
%-----------------------------
\begin{equation}\label{seq.28}
\mathcal{H}^{(3)}_{\mathrm{eff}} = \frac{i}{2}\big[ \bs^{(2)}, \mathcal{H}_1 \big] + \frac{1}{6} \big[ \bs^{(1)}, \big[ \bs^{(1)}, \mathcal{H}_1 \big].
\end{equation}
%-----------------------------
Using the projector-based decomposition, we can continue with the algebraic equations of motion for the elements of the generating function $\bs^{(2)}$. They are given by
%-------------------- 
\begin{subequations}
\begin{align}
\mathcal{P}\bs^{(2)} \mathcal{Q} \mathcal{Q} \mathcal{H}_0 \mathcal{Q} - \cancel{\mathcal{P} \mathcal{H}_0 \mathcal{Q} \mathcal{Q} \bs^{(2)} \mathcal{Q}} & = \cancel{\mathcal{P} \mathcal{H}_{\mathcal{T}} \mathcal{Q} \mathcal{Q} \bs^{(1)} \mathcal{Q}} - \mathcal{P} \bs^{(1)} \mathcal{Q} \mathcal{Q} \mathcal{H}_1 \mathcal{Q} \nonumber \\
\label{seq.29.1}
\Rightarrow \mathcal{P}\bs^{(2)} \mathcal{Q} \mathcal{Q} \mathcal{H}_0 \mathcal{Q} & = - \mathcal{P} \bs^{(1)} \mathcal{Q} \mathcal{Q} \mathcal{H}_1 \mathcal{Q}, \\
\cancel{\mathcal{Q}\bs^{(2)} \mathcal{Q} \mathcal{Q} \mathcal{H}_0 \mathcal{P}} - \mathcal{Q} \mathcal{H}_0 \mathcal{Q} \mathcal{Q} \bs^{(2)} \mathcal{P} & = \mathcal{Q} \mathcal{H}_1 \mathcal{Q} \mathcal{Q} \bs^{(1)} \mathcal{P} - \cancel{\mathcal{Q} \bs^{(1)} \mathcal{Q} \mathcal{Q} \mathcal{H}_1 \mathcal{P}} \nonumber \\
\label{seq.29.2}
\Rightarrow \mathcal{Q} \mathcal{H}_0 \mathcal{Q} \mathcal{Q} \bs^{(2)} \mathcal{P} & = - \mathcal{Q} \mathcal{H}_1 \mathcal{Q} \mathcal{Q} \bs^{(1)} \mathcal{P}.
\end{align}
\end{subequations}
%--------------------

%-----------------------------------------------------------------------------
\subsection{Second-order multipolar exchange interaction \label{sec:sec.2.2}}
%-----------------------------------------------------------------------------

We can now compute the effective low-energy multipolar exchange interaction in edge-sharing octahedral configurations. In the context of octahedral environments, it is essential to consider the anisotropic nature of the Slater-Koster hopping parameters between neighboring sites. For the analysis, we will focus on a specific bond connecting two neighboring TM ions, which we will label as the $z$-bond aligned in the $xy$-plane. Similar analyses can be carried out for the exchange interactions along other bonds by applying the appropriate rotations. First, we consider the perturbative hopping Hamiltonian along the $z$-bond as 
%-----------------------------
\begin{equation}\label{seq.30}
\mathcal{H}_1 
= 
-t_2 \sum_{\langle ij \rangle; \sigma} \left( d^{\dagger}_{i;yz \sigma} d_{j;zx \sigma} + d^{\dagger}_{i;zx \sigma} d_{j;yz \sigma} \right) + \rm{h.c.}, 
\end{equation} 
%----------------------------- 
where $t_2$ is the nearest-neighbor (NN) hopping amplitude. We now follow Eq.~\eqref{seq.23} to write the effective Hamiltonian in the second order as ($\mathcal{P}_{\rm{low}} = \mathcal{P}_i \mathcal{P}_j$ is the product of the low-energy projection operators in two sites)
%-----------------------------
\begin{align}
\nonumber
&
\mathcal{P}_{\rm{low}} \mathcal{H}^{(2)}_{\rm{eff}} \mathcal{P}_{\rm{low}} \\
\label{seq.31}
& = 
-\frac{1}{2\mathcal{Q}_j \mathcal{H}^{(j)}_0 \mathcal{Q}_j} \mathcal{P}_i \mathcal{H}_1 \mathcal{Q}_j \mathcal{Q}_j \mathcal{H}_1 \mathcal{P}_i + i \leftrightarrow j + \rm{h.c.}, \\
\nonumber
& =
-\frac{t_2^2}{2} \sum_{\substack{\{n_l,\eta_l\} \\ \alpha\beta}} 
\frac{\bra{i,n_1; j,n_2} d^{\dag}_{i\beta\sigma} d_{j\alpha \sigma} \ket{\psi_{i,\eta_1}; \phi_{j,\eta_2}}
\bra{\psi_{i,\eta_1}; \phi_{j,\eta_2}} d^{\dag}_{j\alpha\sigma} d_{i\beta \sigma} \ket{i,n_3; j,n_4}}{\Delta E_{\rm{ex}}^{\{n_i,\eta_i\}}}
\ket{i,n_1; j,n_2} \bra{i,n_3; j,n_4} + i \leftrightarrow j + \rm{h.c.},
\end{align}
%-----------------------------
where $\ket{i,n} \forall n = \Uparrow, \Downarrow$ corresponds to the ground-state doublets as mentioned in Eq.~\eqref{seq.15}, $\ket{\psi_{i,\eta}} \forall \eta = \rm{A,B,C,D}$ and $\ket{\phi_{i,\eta}} \forall \eta = \rm{A,B,C,D}$ correspond to the three-electron and one-electron intermediate states, respectively. Finally, $\Delta E_{\rm{ex}}^{\{n_i,\eta_i\}}$ corresponds to the atomic excitation energy for the virtual process involved. It is computed as~\cite{PhysRevB.95.014409}
%-----------------------------
\begin{equation}\label{seq.32}
\Delta E_{\rm{ex}}^{\{n_i,\eta_i\}}
= 
E^{(i)} (N_i = 3) + E^{(j)} (N_j = 1) - E^{(i)}(N_i = 2) - E^{(j)} (N_j = 2) = U - 3J_{\rm{H}} + \lambda.
\end{equation}
%-----------------------------
Note that the virtual hopping between the TM atoms changes the electron filling in the intermediate states as 
%-----------------------------
\begin{equation}\label{seq.33}
d_i^2 d_j^2 \rightarrow d_i^3 d_j^1 \rightarrow d_i^2 d_j^2 + i \leftrightarrow j.
\end{equation}
%-----------------------------
For latter reference, we provide the expressions for the intermediate one- and three-electron ground states within the octahedral environment, ordered by their hierarchical energy distribution as previously outlined. These states are defined as follows~\cite{Georges2013,PhysRevB.95.014409}
%------------------
\begin{subequations}
\begin{align}
\label{seq.34.1}
\ket{\psi_{i,{\rm{A}}}} 
& 
= 
d^{\dag}_{i;xy\uparrow} d^{\dag}_{i;yz\uparrow} d^{\dag}_{i;zx\uparrow} \ket{0}, \\
\label{seq.34.2}
\ket{\psi_{i,{\rm{B}}}} 
&
=
d^{\dag}_{i;xy\downarrow} d^{\dag}_{i;yz\downarrow} d^{\dag}_{i;zx\downarrow} \ket{0}, \\
\label{seq.34.3}
\ket{\psi_{i,{\rm{C}}}} 
& 
= 
\frac{1}{\sqrt{3}}
\left(
d^{\dag}_{i;xy\uparrow} d^{\dag}_{i;yz\uparrow} d^{\dag}_{i;zx\downarrow} +
d^{\dag}_{i;xy\uparrow} d^{\dag}_{i;yz\downarrow} d^{\dag}_{i;zx\uparrow} + 
d^{\dag}_{i;xy\downarrow} d^{\dag}_{i;yz\uparrow} d^{\dag}_{i;zx\uparrow}
\right) 
\ket{0}, \\
\label{seq.34.4}
\ket{\psi_{i,{\rm{D}}}} 
&
= 
\frac{1}{\sqrt{3}}
\left(
d^{\dag}_{i;xy\downarrow} d^{\dag}_{i;yz\downarrow} d^{\dag}_{i;zx\uparrow} +
d^{\dag}_{i;xy\downarrow} d^{\dag}_{i;yz\uparrow} d^{\dag}_{i;zx\downarrow} + 
d^{\dag}_{i;xy\uparrow} d^{\dag}_{i;yz\downarrow} d^{\dag}_{i;zx\downarrow}
\right) 
\ket{0}), \\
\label{seq.34.5}
\ket{\phi_{i,{\rm{A}}}} 
& 
= 
\frac{1}{\sqrt{2}}
\left( 
d^{\dag}_{i;yz,\uparrow} + i d^{\dag}_{i;zx,\uparrow} 
\right) 
\ket{0}, \\
\label{seq.34.6}
\ket{\phi_{i,{\rm{B}}}} 
& 
= 
\frac{1}{\sqrt{2}}
\left( 
d^{\dag}_{i;yz,\downarrow} - i d^{\dag}_{i;zx,\downarrow} 
\right) 
\ket{0}, \\
\label{seq.34.7}
\ket{\phi_{i,{\rm{C}}}} 
& 
= 
\sqrt{\frac{2}{3}} 
\left( 
d^{\dag}_{i;xy,\uparrow} - \frac{d^{\dag}_{i;yz,\downarrow} + i d^{\dag}_{i;zx,\downarrow}}{2} \right)
\ket{0}, \\
\label{seq.34.8}
\ket{\phi_{i,{\rm{D}}}} 
& 
= 
\sqrt{\frac{2}{3}} 
\left( 
d^{\dag}_{i;xy,\downarrow} + \frac{d^{\dag}_{i;yz,\uparrow} - i d^{\dag}_{i;zx,\uparrow}}{2} 
\right) 
\ket{0} \,, 
\end{align}
\end{subequations}
%--------------------
where $\ket{0}$ corresponds to the empty $t_{2\dg}$ states. We do not consider the $e_{\dg}$ states explicitly, assuming the crystal field remains large in the intermediate states. In the case of $d^3$-electronic configuration, there are three electrons in $t_{2\dg}$, and in the presence of Hund's coupling, the total orbital angular momentum is completely quenched. Consequently, the SOC is effectively absent in this case. We define new composite operators $b^{\dag}_{i\tilde{\sigma}} \forall \tilde{\sigma} = \uparrow,\downarrow$ which create two non-Kramers states in Eq.~\eqref{seq.15} as
%-----------------------------
\begin{equation}\label{seq.35}
b^{\dag}_{i\uparrow} \ket{0} = \ket{i,\Uparrow}, \quad
b^{\dag}_{i\downarrow} \ket{0} = \ket{i,\Downarrow}, \quad 
\end{equation}
%-----------------------------
and introduce the pseudo-SU(2) generating functions with them as 
%-----------------------------
\begin{equation}\label{seq.36}
\tilde{\bm{\sigma}}_i = b^{\dag}_{i\alpha} {\bm{\tau}}_{\alpha \beta} b_{i\beta},
\end{equation}
%-----------------------------
where $\bm{\tau} = (\tau_x, \tau_y, \tau_z)$ is the vector of Pauli matrices. In this way, we compute all the matrix elements in Eq.~\eqref{seq.30} using \texttt{DiracQ} package in \texttt{MATHEMATICA V.14.1}~\cite{Shastry2013}, and subsequently rewrite the resulting expressions in the pseudo-spin operators $(\tilde{\sigma}^x, \tilde{\sigma}^y, \tilde{\sigma}^z)$.

%-----------------------------------------------
\begin{table}[t]
\caption{The summary of the transformation of various quantities under the irreducible representations of the dihedral group $D_{4h}$~\cite{PhysRevB.106.155155}. This transformation applies to the edge-sharing geometry.}\label{tab:symmetry}
\centering
\begin{tabular}{|c |c |c |c |}
\hline
Object	&	$\quad S_{4z} \quad$	&	$\quad C_3 \quad$	&	$\quad \mathcal{T} \quad$ \\
\hline \hline
$x$		 			&	$-y$		&	$y$		&	$x$ \\	
\hline
$y$		 			&	$x$			&	$z$		&	$y$ \\	
\hline
$z$		 			&	$-z$		&	$x$		&	$z$ \\	
\hline
$\tilde{\sigma^x}$	&	$\tilde{\sigma^x}$		&	$-\frac{1}{2} \tilde{\sigma^x} - \frac{\sqrt{3}}{2} \tilde{\sigma^z}$		&	$\tilde{\sigma^x}$ \\	
\hline
$\tilde{\sigma^y}$ 	&	$-\tilde{\sigma^y}$		&	$\tilde{\sigma^y}$		&	$-\tilde{\sigma^y}$ \\	
\hline
$\tilde{\sigma^z}$ 	&	$-\tilde{\sigma^z}$		&	$ \frac{\sqrt{3}}{2}\tilde{\sigma^x} - \frac{1}{2}\tilde{\sigma^z}$		&	$\tilde{\sigma^z}$ \\	
\hline
\end{tabular}
\end{table}
%-----------------------------------------------

In this context, the hopping mechanism is characterized by non-conservation of orbital properties, as indicated in Eq.~\eqref{seq.30}. Before proceeding further, we explore the transformations of multipolar operators under the dihedral group generators, specifically $S_{4z}$ and $C_3$ rotations around the (111) axis for the edge-sharing geometry. The corresponding transformations are outlined in Table~\ref{tab:symmetry}. Adhering to these symmetry constraints, we derive a second-order Hamiltonian that aligns with previous theoretical studies~\cite{PhysRevLett.111.197201,PhysRevB.107.L020408,PhysRevResearch.3.033163,PhysRevB.105.014438}. In the absence of Hund's coupling and spin-orbit coupling (SOC) effects in the intermediate states, and considering all 120 intermediate states ($^6C_1$ $\times$ $^6C_3$), resulting from combinations of $d^1$ and $d^3$ electronic configurations, the effective low-energy multipolar exchange Hamiltonian exhibits a distinctive bond-independent character as
%---------------------
\begin{subequations}
\begin{align}
\label{seq.37.1}
\mathcal{H}^{(0)}_{\rm{eff}} & = \mathcal{H}^{(0)}_{x} + \mathcal{H}^{(0)}_{y} + \mathcal{H}^{(0)}_{z}, \\
\label{seq.37.2}
\mathcal{H}^{(0)}_{\eta} & = \frac{2t_2^2}{3 \Delta E_{\rm{ex.}}}
\sum_{\langle ij \rangle, z} 
\left(  
\tilde{\sigma}^y_{i} \tilde{\sigma}^y_{j} 
- 
\tilde{\sigma}^x_{i} \tilde{\sigma}^x_{j} - 
\tilde{\sigma}^z_{i} \tilde{\sigma}^z_{j} 
\right), \quad \eta \in \{x,y,z\}.
\end{align}
\end{subequations}
%---------------------
The multipolar exchange Hamiltonian takes an isotropic and bond-independent form, as articulated in Eqs.~\eqref{seq.37.1}-\eqref{seq.37.2}. The bilinear exchange form closely resembles the bond-dependent Kugel-Khomskii orbital exchange model~\cite{Kugel_1982}. We further note that the cubic rotation symmetry along the $[111]$ direction constrains the quadrupolar exchange interaction to be invariant under $C^3_{[111]}$ rotation (see Table.~\ref{tab:symmetry}). Note that for $x$ and $y$-bonds the pseudo-spin up and down states in Eq.~\eqref{seq.16.1}, and Eq.~\eqref{seq.16.2} are rotated accordingly so that the transformed $\tilde{\sigma}^z$ is along the projected octahedral $x$ and $y$-axis on the $[111]$ plane.

%-----------------------------------------------------------------------------
\subsection{Third-order multipolar exchange interaction \label{sec:sec.2.3}}
%-----------------------------------------------------------------------------

\begin{figure}[t!]
\centering \includegraphics[width=0.6\linewidth]{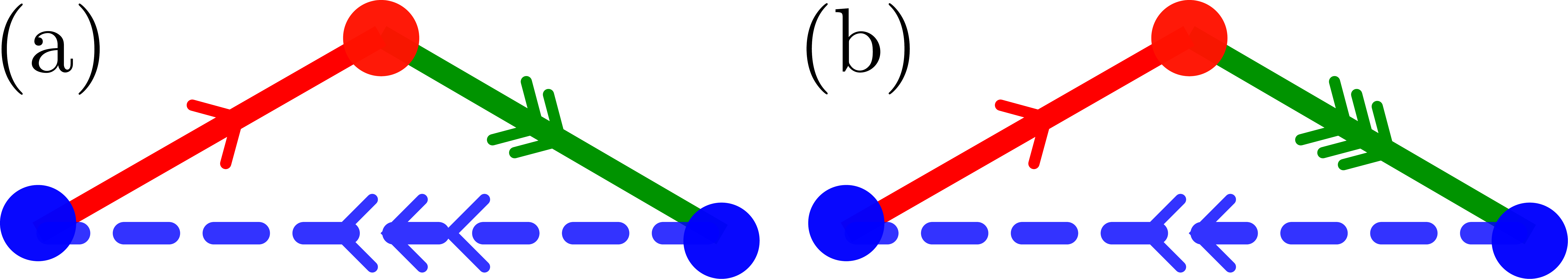}
\caption{Two types of virtual exchange of electrons in the third-order perturbation expansion of the effective Hamiltonian derivation in Eq.~\eqref{seq.38}. (a) a direct hopping process as illustrated by the arrows, and (b) an indirect hopping in the intermediate states labeled by the arrow. The number of arrows dictates the step of the virtual hopping in the third-order super-exchange process.}\label{fig:SFig2}
\end{figure}

In this section, we derive the third-order expression for the low-energy effective Hamiltonian between the multipolar degrees of freedom. Utilizing the dynamical equations for the generator $\bs^{(2)}$ in Eqs.~\eqref{seq.29.1} and ~\eqref{seq.29.2}, we obtain the effective Hamiltonian from Eq.~\eqref{seq.28} as
%-------------
\begin{align}
\label{seq.38}
\mathcal{H}^{(3)}_{\rm{eff}} = &  \frac{2t_2^2 t_2'}{3\Delta E_{\rm{ex.}}^2} \sin \frac{e\Phi}{\hbar c}
\sum_{\substack{ \llangle ijk \rrangle \\ \bigtriangleup}}
\Bigg[\left(
\frac{\tilde{\sigma}^y_{i}\tilde{\sigma}^y_{j}\tilde{\sigma}^y_{k}}{3} -
\frac{2\tilde{\sigma}^z_{i}\tilde{\sigma}^z_{j}\tilde{\sigma}^y_{k}}{9} -
\frac{2\tilde{\sigma}^y_{i}\tilde{\sigma}^z_{j}\tilde{\sigma}^z_{k}}{9} -
\frac{2\tilde{\sigma}^z_{i}\tilde{\sigma}^y_{j}\tilde{\sigma}^z_{k}}{9} -
\frac{2\tilde{\sigma}^x_{i}\tilde{\sigma}^x_{j}\tilde{\sigma}^y_{k}}{9} -
\frac{2\tilde{\sigma}^y_{i}\tilde{\sigma}^x_{j}\tilde{\sigma}^x_{k}}{9} -
\frac{2\tilde{\sigma}^x_{i}\tilde{\sigma}^y_{j}\tilde{\sigma}^x_{k}}{9} \right) \Bigg].
\end{align}
%-------------
The overlap matrix elements are again computed in \texttt{DiracQ} package in \texttt{MATHEMATICA V.14.1}~\cite{Shastry2013}. The trilinear term is only active in the presence of an external magnetic field $B$. Here, $\Phi$ is the total flux enclosed within the triangular plaquette in the presence of $B$. This is consistent with the $C_3$ rotations and overall time-reversal symmetry of the system. We incorporate Eq.~\eqref{seq.28} to derive the Hamiltonian above. Utilizing the generating functions, the virtual hopping processes can be classified into two types as illustrated in Fig.~\ref{fig:SFig2}(a,b) \--- among which only process (a) contributes to the form in Eq.~\eqref{seq.38}. The contribution from the process in panel (b) of Fig.~\ref{fig:SFig2} identically vanishes due to the conservation of $J_z$ angular momentum. In the above derivation the states for three different bond types enter as illustrated in Fig.~\ref{fig:SFig2}. The pseudo-spin up and pseudo-spin down operators for each bond are rotated in the usual manner as for the $\rm{E}_\dg$ doublet as discussed in the main text.

%-------------------------------------------------------------------------------------------
%-------------------------------------------------------------------------------------------

%------------------------------------------------------
\section{Multipolar multiferroicity \label{sec:sec.4}}
%------------------------------------------------------

In this section, we provide the details of the derivation leading to Eq.~(6) and Eqs.~(7a)-(7b) in the main text for the ferroelectric polarization and the electrical quadrupolar responses. Motivated by our MC simulation leading to a non-collinear multipolar texture obtained by the exchange Hamiltonian in Eqs.~\eqref{seq.37.1}-\eqref{seq.38}, we consider the local approximation of the Hamiltonian as discussed in the main text. The corresponding Hamiltonian is written as 
%------------------------------
\begin{equation}\label{seq.39}
\mathcal{H}_0 = - \tilde{U} \sum_i \bm{e}_i \cdot \bm{\tilde{\sigma}}_i,
\end{equation}
%------------------------------
where $\bm{e}_i = (\sin\theta_i \cos \phi_i, \sin \theta_i \sin \phi_i, \cos \theta_i)$, $\bm{\tilde{\sigma}}_i$ corresponds to the local pseudo-spin at site $i$, and $\tilde{U}$ corresponds to a phenomenological energy scale (associated with the Hubbard interaction, Hund's coupling strength, SOC and the local crystal field). Now, we rewrite the $\ket{\rm{E}_\dg}$ states in terms of the two $d$-orbital operators for simplicity as
%---------------------
\begin{subequations}
\begin{align}
\label{seq.40.1}
\ket{i;\Uparrow} & = 
\left( 
\frac{d^{\dag}_{i;xy\uparrow} d^{\dag}_{i;zx\uparrow}}{2} + \frac{d^{\dag}_{i;xy\downarrow} d^{\dag}_{i;zx\downarrow}}{2} + 
\frac{i d^{\dag}_{i;xy\uparrow} d^{\dag}_{i;yz\uparrow}}{2} - \frac{ i d^{\dag}_{i;xy\downarrow} d^{\dag}_{i;yz\downarrow}}{2}
\right) \ket{0}, \\
\label{seq.40.2}
\ket{i;\Downarrow} & = 
\left(
\frac{d^{\dag}_{i;xy\downarrow} d^{\dag}_{i;zx\downarrow}}{2\sqrt{3}} + \frac{d^{\dag}_{i;xy\uparrow} d^{\dag}_{i;zx\uparrow}}{2\sqrt{3}} + 
\frac{d^{\dag}_{i;yz\uparrow} d^{\dag}_{i;zx\downarrow}}{\sqrt{3}} + \frac{d^{\dag}_{i;yz\downarrow} d^{\dag}_{i;zx\uparrow}}{\sqrt{3}} + 
\frac{i d^{\dag}_{i;xy\downarrow} d^{\dag}_{i;yz\downarrow}}{2\sqrt{3}} - \frac{i d^{\dag}_{i;xy\uparrow} d^{\dag}_{i;yz\uparrow}}{2\sqrt{3}} 
\right) \ket{0},
\end{align}
\end{subequations}
%---------------------
where $\ket{0}$ is the empty $t_{2\dg}$-sector. Adopting the Pauli matrix structure, the Hamiltonian in Eq.~\eqref{seq.39} can be diagonalized to obtain the eigenfunctions as 
%----------------------
\begin{subequations}
\begin{align}
\label{seq.41.1}
\ket{\psi_{i}^+} & = \sin \frac{\theta_i}{2} \ket{i; \Uparrow} + e^{i\phi_i} \cos \frac{\theta_i}{2} \ket{i; \Downarrow}, \\
\label{seq.41.2}
\ket{\psi_{i}^-} & = \cos \frac{\theta_i}{2} \ket{i; \Uparrow} - e^{i\phi_i} \sin \frac{\theta_i}{2} \ket{i; \Downarrow}.
\end{align}
\end{subequations}
%----------------------
Here, $\ket{\psi^+}$ is the ground state with energy $-\tilde{U}$, and $\ket{\psi^-}$ is the excited state with energy $+\tilde{U}$. Considering the ground state as $\ket{\psi^+}$, we now consider two different ligand-TM hopping processes to derive the multipolar charge responses. The ligand-TM hopping is written as
%----------------------------
\begin{equation}\label{seq.42}
\mathcal{H}_{\rm{hop}} 
= 
t^{\pi}_{pd} d^{\dag}_{i;zx\sigma} p_{l';z\sigma}
+
t^{\pi}_{pd} d^{\dag}_{i;yz\sigma} p_{l;z\sigma}
-
t^{\pi}_{pd} d^{\dag}_{j;zx\sigma} p_{l';z\sigma}
-
t^{\pi}_{pd} d^{\dag}_{j;yz\sigma} p_{l;z\sigma}
+
\rm{h.c.},
\end{equation}
%----------------------------
where $l$ ($l'$) corresponds to the upper (lower) ligand sites in Fig.~\ref{fig:SFig3}(a,b).

%-----------------------------------------------------------
\subsection{Ferroelectric polarization \label{sec:sec.4.1}}
%-----------------------------------------------------------

\begin{figure}[t!]
\centering \includegraphics[width=0.8\linewidth]{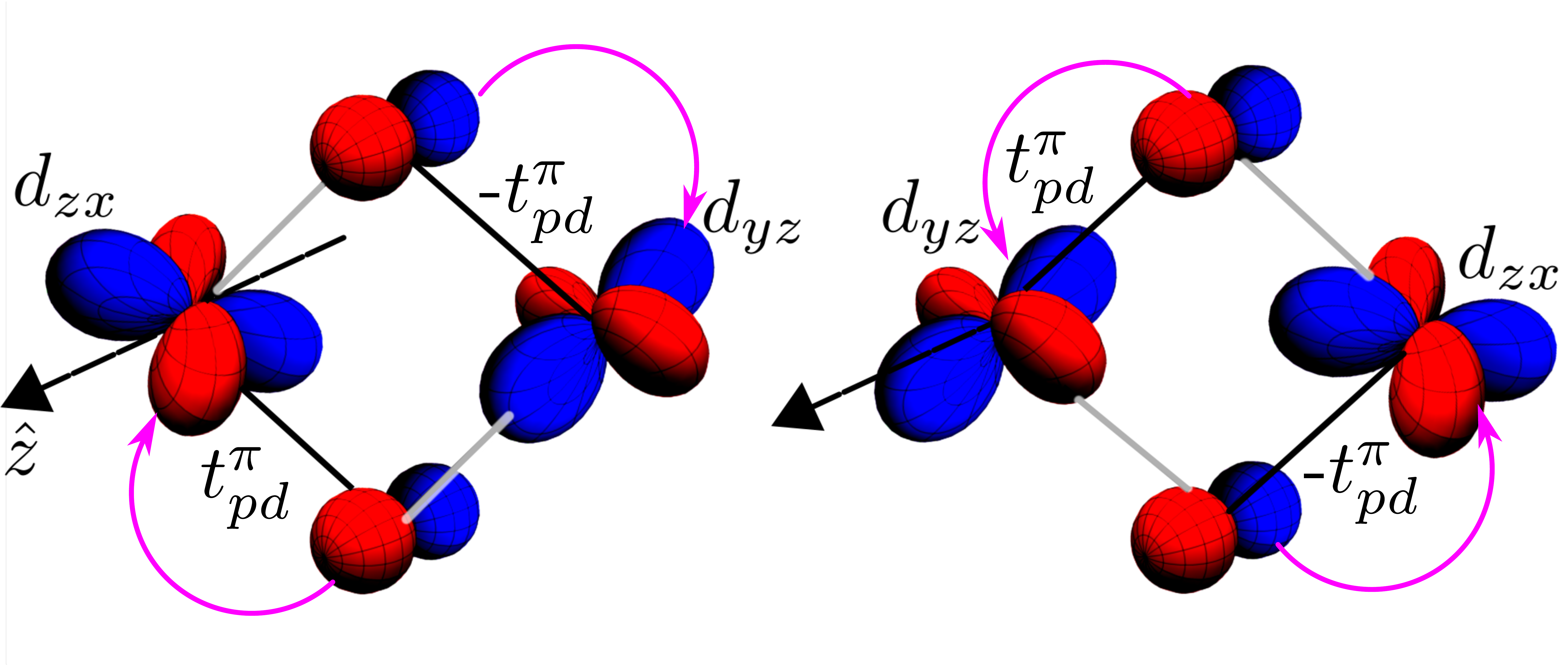}
\caption{(a) The virtual hopping of electrons from a \SB{\textbf{filled}} ligand site to a TM site along the $z$-bond orientation in the edge-shared geometry. The hopping amplitudes differ in sign between the upper and lower triangles in the TM-ligand-TM clusters. (b) The virtual hopping of electrons to an \SB{\textbf{empty}} ligand site from a TM site along the $z$-bond orientation in the edge-shared geometry. The hopping amplitudes differ in sign between the left and right triangles in the TM-ligand-TM clusters.}\label{fig:SFig3}
\end{figure}

We consider a four-site cluster with two TM and two ligand sites. In the stoichiometric configuration of the TM compound, we consider filled ligand sites with electrons. For subsequent analysis, we rewrite Eq.~\eqref{seq.41.1} in the following manner as
%-----------------------------
\begin{equation}\label{seq.43}
\ket{\psi_i^+} = 
c_i \underbrace{(\ket{xy\uparrow}_i\ket{zx\uparrow}_i + \ket{xy\downarrow}_i\ket{zx\downarrow}_i)}_{yz}
+
id_i \underbrace{(\ket{xy\uparrow}_i\ket{yz\uparrow}_i - \ket{xy\downarrow}_i\ket{yz\downarrow}_i)}_{zx}
+
a_i \underbrace{(\ket{yz\uparrow}_i\ket{zx\downarrow}_i + \ket{yz\downarrow}_i\ket{zx\uparrow}_i)}_{xy},
\end{equation}
%-----------------------------
where 
$$
c_i = \frac{\sqrt{3} \sin \frac{\theta_i}{2} + e^{i\phi_i} \cos \frac{\theta_i}{2}}{2\sqrt{3}}, \quad
d_i = \frac{\sqrt{3} \sin \frac{\theta_i}{2} - e^{i\phi_i} \cos \frac{\theta_i}{2}}{2\sqrt{3}}, \quad
a_i = c_i - d_i,
$$ 
and we underlined the effective spatial structure of the two-electron composite states in Eq.~\eqref{seq.43}. We now consider virtual charge transfer between the ligand and the TM atoms within a double exchange framework. The associated paths are shown in Fig.~\ref{fig:SFig3}(a). \SB{This process leads to an intermediate state with a stoichiometric configuration \--- \(d^3\) and \(p_z^1\). Additionally, we take into account the strong Hund's coupling arising from the heavy TM atom. In the case of a $d^3$ configuration, where the $t_{2\dg}$ orbitals are half-filled, the spin-orbit coupling is inactive as the orbital angular momentum is fully quenched in the presence of strong Hund's coupling. The corresponding low-energy atomic ground states for a $d^3$ ion in an octahedral crystal field, in the presence of strong Hund's coupling, are provided previously in Eqs.~\eqref{seq.34.1} to \eqref{seq.34.4}. To calculate the correction to the ground states presented in Eq.~\eqref{seq.41.1}, we utilized these states within a four-site cluster illustrated in Fig.~\ref{fig:SFig3}(a) and performed first-order perturbation theory to obtain}
%-------------
\begin{align}
\ket{\psi_{\rm{dip.}}} = 
\nonumber
\frac{1}{N} \bigg[ 
\ket{\psi^+_i} \otimes \ket{p^2_z}_{l} + \ket{\psi^+_j} \otimes \ket{p^2_z}_{l'} + \frac{t^{\pi}_{pd}}{\Delta_{\rm{crys.}}}
& id_i \left( 
\ket{xy\uparrow}_i\ket{yz\uparrow}_i\ket{zx\sigma}_i \ket{p_z \bar{\sigma}}_{l'}
- 
\ket{xy\downarrow}_i\ket{yz\downarrow}_i\ket{zx\sigma}_i \ket{p_z \bar{\sigma}}_{l'}
\right) \\
\nonumber
-
\frac{t^{\pi}_{pd}}{\Delta_{\rm{crys.}}}
& id_j \left( 
\ket{xy\uparrow}_j\ket{yz\uparrow}_j\ket{zx\sigma}_j \ket{p_z \bar{\sigma}}_l
- 
\ket{xy\downarrow}_j\ket{yz\downarrow}_j\ket{zx\sigma}_j \ket{p_z \bar{\sigma}}_l
\right) \\ \nonumber
+
\frac{t^{\pi}_{pd}}{\Delta_{\rm{crys.}}} 
&
c_i  \left( 
\ket{xy\uparrow}_i\ket{zx\uparrow}_i\ket{yz\sigma}_i \ket{p_z \bar{\sigma}}_l
+
\ket{xy\downarrow}_i\ket{zx\downarrow}_i\ket{yz\sigma}_i \ket{p_z \bar{\sigma}}_l
\right) \\
\label{seq.44}
-
\frac{t^{\pi}_{pd}}{\Delta_{\rm{crys.}}}
&
c_j  \left( 
\ket{xy\uparrow}_j\ket{zx\uparrow}_j\ket{yz\sigma}_j \ket{p_z \bar{\sigma}}_{l'}
+
\ket{xy\downarrow}_j\ket{zx\downarrow}_j\ket{yz\sigma}_j \ket{p_z \bar{\sigma}}_{l'}
\right) \bigg], 
\end{align}
%-------------
where $N$ is the normalization factor, $\Delta_{\rm{crys.}}$ is the charge-transfer energy between TM $d$ and ligand $p$ orbitals, $l$ ($l'$) corresponds to the upper (lower) ligand sites in Fig.~\ref{fig:SFig3}(a), and $\ket{p_z^2}$ corresponds to a filled ligand site with two electrons. \SB{Note that we have used the notation $\sigma,\bar{\sigma}$ for the spin configurations on the ligand and the TM sites. In the stable stoichiometric configuration of the underlying material, typically the ligand $p_z$ orbital is fully quneched in spins as it has both up and down components. However, because of the hopping of the electrons from the ligand to the TM atom through the Hamiltonian in Eq.~\eqref{seq.42} the spin configuration between the remaining ligand and TM ion will be anti-parallel.}

%-------------------------------------------------------------------------------------------
%-------------------------------------------------------------------------------------------

Now, in the perturbed state $\ket{\psi}$, we compute the expectation value of the polarization operator ${\bf{P}} = e \dr$. Using the parity of the associated states in Eq.~\eqref{seq.41.1} and Eq.~\eqref{seq.43}, \SB{it can be shown by a straightforward but tedious algebra that only} $x$-component of the polarization survives upon evaluating the overlap integrals in $\braket{\psi_{\rm{dip.}}|\dr|\psi_{\rm{dip.}}}$. \SB{See the discussion below Eq.~(6) in the main text, for the parity based argument for vanishing of certain types of integrals. For algebraic simplication, we utilized the DiracQ package to compute such overlaps between the states. We keep terms only up to the first order in $t^\pi_{pd}$ i.e., ignore $\mathcal{O}((t^\pi_{pd})^2)$ contributions.} The polarization for the other bonds can be analogously obtained by using appropriate perturbed states upon different ligand-TM hoppings. \SB{The induced ferroelectric polarization can be computed in a crude manner by considering the charge fluctuations between the relevant orbitals as $I = \int d^3\dr \, d_{zx}(\dr) x p_{z}(\dr)$ for $z$-bond. For other bonds, one can utilize the cubic rotation symmetry and permute the coordinates to obtain the result in Eq.~(6) in the main text.}

%--------------------------------------------------------------------
\subsection{Electrical quadrupolar distribution \label{sec:sec.4.2}}
%--------------------------------------------------------------------

Here, we provide the \SB{detailed derivation} for the \SB{two bond-dependent} electrical quadrupole \SB{moments in Eqs.~(7a) and (7b)} in the main text. In this case, we consider a similar \SB{geometry of the four-site cluster}, but focus on those \SB{TM compounds where in the stable crytallographic configuration, the ligand sites are empty in the electron picture. We subsequently consider a double-exchange cotunneling process \--- one electron in the TM atom within the $d^2$ configuration tunnels into the upper ligand ion, while the remaining one hops into the lower ligand ion as illustrated in Fig.~\ref{fig:SFig3}(b). Similar to the double-exchange path in Fig.~\ref{fig:SFig3}(b), here, we also consider both the left and the right cotunnelings in the perturbation theory.} The perturbation calculations are performed exactly similar to the polarization in the previous section. \SB{However, a subtle distinction arises from the previous case in the context of the relative sign between the left and the right perturbation process. Since, the cotunneling amplitude is quadratic in $t^\pi_{pd}$ we do obtain any relative sign in the perturbed eigenstates. Again using a first-order perturbation theory, we obtain}
%-----------------------------
\begin{equation}\label{seq.45}
\ket{\psi_{\rm{quad.}}} = 
\frac{1}{N} \bigg[ 
\ket{\psi^+_i} + \ket{\psi^+_j} + 
\left(\frac{t^{\pi}_{pd}}{\Delta_{\rm{crys.}}}\right)^2
(2a_i + 2a_j) \left( 
\ket{p_z\uparrow}_{l} \ket{p_z\downarrow}_{l'}
+
\ket{p_z\downarrow}_{l} \ket{p_z\uparrow}_{l'}
\right)\bigg] \,.      
\end{equation}
%-----------------------------
\SB{Note the relative sign between $a_i$ and $a_j$ is positive. Next, we consider the electrical quadrupolar moments in the tesseral harmonic notation as $\mathcal{Q}^{\rm{s}}_{21} \sim zx$, and $\mathcal{Q}^{\rm{s}}_{22} \sim yz$~\cite{Kusunose2008,Hayami2018}. See Eq.~(1) in the main text for definition. Finally, we compute the expectation values of $\mathcal{Q}_{lm}$ in the perturbed eigenstate as $\braket{\psi_{\rm{quad.}}|\mathcal{Q}_{lm}|\psi_{{\rm{quad.}}}}$. The calculation is much simpler in this case. Through a straightforward algebra, it can be shown for $z$-bond, only $\mathcal{Q}^{\rm{s}}_{21}$, and $\mathcal{Q}^{\rm{s}}_{22}$ expectation values are non-vanishing and all the other components vanish. One may utilize the cubic rotation symmetry to compute the non-zero quadrupole moments for the other bonds as before. An estimate for the induced ferroquadrupolar moment can be similarly considered by only considering the charge fluctuations in between the relevant orbitals as explained in Eqs.~(7a,7b) in the main text. In a crude approximation, the estimate for the overlap integrals in $\braket{\psi_{\rm{quad.}}|\mathcal{Q}_{lm}|\psi_{{\rm{quad.}}}}$ can be computed in a similar fashion to the earlier case. We notice that there will be now two overlap integrals between the two $d$, and $p$-orbitals. Roughly, these two integrals can be evaluated independently and is approximately equal to the square of the overlap integral $I = \int d^3\dr \, d_{zx}(\dr) x p_{z}(\dr)$ for the polarization estimate, \textit{i.e.} $I' \approx I^2$.}

\end{document}